\documentclass[10pt,article,apl,aps,prb,superscriptaddress,amsmath,amssymb,floatfix,twocolumn,tightenlines]{revtex4-2}

\usepackage{textcomp}
\usepackage{setspace}
\usepackage{amsmath}
\usepackage[margin=0.75in]{geometry}
\usepackage{physics}
\usepackage{amsfonts}
\usepackage{amssymb}
\usepackage{bm}
\usepackage{MnSymbol}
\usepackage{graphicx}
\usepackage{natbib}
\usepackage{gensymb}
\usepackage{setspace}
\usepackage{bbm}
\usepackage{tabularray}
\UseTblrLibrary{booktabs}

\usepackage{xspace}
\usepackage{graphicx}
\usepackage[dvipsnames]{xcolor}
\usepackage[separate-uncertainty=true,multi-part-units=single,print-unity-mantissa=false]{siunitx}
\usepackage[version=4]{mhchem}
\usepackage{xfp}
\newcommand\ExtendedData{%
    \xdef\preextfigures{\arabic{figure}}
    \renewcommand\thefigure{\textbf{\fpeval{\arabic{figure}-\preextfigures}}}
    \renewcommand{\figurename}{\textbf{\extlbl}}
    \xdef\preexttables{\arabic{table}}
    \renewcommand\thetable{\textbf{\fpeval{\arabic{table}-\preexttables}}}
    \renewcommand{\tablename}{\textbf{\exttbllbl}}
}

\newcommand{\figlbl}{Fig.}
\newcommand{\figlbls}{Figs.}
\newcommand{\extlbl}{Extended~Data~Fig.} 
\newcommand{\extlbls}{Extended~Data~Figs.} 
\newcommand{\exttbllbl}{Extended~Data~Table} 

\renewcommand{\figurename}{\textbf{\figlbl}} 
\renewcommand{\thefigure}{\textbf{\arabic{figure}}}
\newcommand{\figtitle}[1]{\textbf{#1}\xspace} 

\newcommand{\panel}[1]{\textbf{#1}\xspace}
\newcommand{\moire}{moir{\'e}\xspace}
\newcommand{\Moire}{Moir{\'e}\xspace}
\newcommand{\supermoire}{supermoir{\'e}\xspace}
\newcommand{\Supermoire}{Supermoir{\'e}\xspace}

\usepackage[colorlinks, citecolor={blue}, linkcolor={black}, urlcolor={blue}]{hyperref}

\begin{document}

\title{Magic continuum in multi-\moire twisted trilayer graphene}

\author{Li-Qiao Xia}
\thanks{These authors contributed equally.}
\email{xialq@mit.edu}
\affiliation{Department of Physics, Massachusetts Institute of Technology, Cambridge, Massachusetts 02139, USA}

\author{Aviram Uri}
\thanks{These authors contributed equally.}
\affiliation{Department of Physics, Massachusetts Institute of Technology, Cambridge, Massachusetts 02139, USA}

\author{Jiaojie Yan}
\thanks{These authors contributed equally.}
\affiliation{Max Planck Institute for Solid State Research, 70659, Stuttgart, Germany}

\author{Aaron Sharpe}
\thanks{These authors contributed equally.}
\affiliation{Department of Physics, Stanford University, Stanford, California 94305, USA}
\affiliation{Stanford Institute for Materials and Energy Sciences, SLAC National Accelerator Laboratory, Menlo Park, California 94025, USA}

\author{Filippo Gaggioli}
\affiliation{Department of Physics, Massachusetts Institute of Technology, Cambridge, Massachusetts 02139, USA}

\author{Nicole S. Ticea}
\affiliation{Department of Applied Physics, Stanford University, Stanford, California 94305, USA}

\author{Julian May-Mann}
\affiliation{Department of Physics, Stanford University, Stanford, California 94305, USA}

\author{Kenji Watanabe}
\affiliation{Research Center for Electronic and Optical Materials, National Institute for Materials Science, 1-1 Namiki, Tsukuba 305-0044, Japan}

\author{Takashi Taniguchi}
\affiliation{Research Center for Materials Nanoarchitectonics, National Institute for Materials Science,  1-1 Namiki, Tsukuba 305-0044, Japan}

\author{Liang Fu}
\affiliation{Department of Physics, Massachusetts Institute of Technology, Cambridge, Massachusetts 02139, USA}

\author{Trithep Devakul}
\affiliation{Department of Physics, Stanford University, Stanford, California 94305, USA}
 
\author{Jurgen H. Smet}
\affiliation{Max Planck Institute for Solid State Research, 70659, Stuttgart, Germany}

\author{Pablo Jarillo-Herrero}
\email{pjarillo@mit.edu}
\affiliation{Department of Physics, Massachusetts Institute of Technology, Cambridge, Massachusetts 02139, USA}

\date{{\small \today}}

\begin{abstract}

\Moire lattices provide a highly tunable platform for exploring the interplay between electronic correlations and band topology~\cite{Andrei2021The}.
Introducing a second \moire pattern extends this paradigm: interference between the two \moire patterns produces a \supermoire modulation, opening a route to further tailor electronic properties.
Twisted trilayer graphene generally exemplifies such a system: two distinct \moire patterns arise from the relative twists between adjacent graphene layers.
Here, we report the observation of correlated phenomena across a wide range of twisted trilayer graphene devices whose twist angles lie along two continuous lines in the twist-angle parameter space~\cite{Zhu2020Twisted,Popov2023Magic,Yang2024Multi,Foo2023Extended}.
Depending on the degree of lattice relaxation, twisted trilayer graphene falls into two classes~\cite{Yang2024Multi}: \moire polycrystals~\cite{Xia2025Topological,Hoke2024Imaging}, composed of periodic domains with locally commensurate \moire order, and \moire quasicrystals, characterized by smoothly varying local \moire configurations~\cite{Uri2023Superconductivity}.
In helically twisted \moire polycrystals, we observe an anomalous Hall effect, consistent with topological bands arising from domains with broken $xy$-inversion symmetry.
In contrast, superconductivity appears generically in our \moire quasicrystals.
A subset of these systems exhibits signatures of spatially modulated superconductivity, which we attribute to the \supermoire structure.
Our findings uncover the organizing principles of the observed correlated phases in twisted trilayer graphene, highlight the critical roles of the \supermoire modulation and lattice relaxation, and suggest a broader framework in which magic conditions arise not as isolated points but as extended manifolds within the multi-dimensional twist-angle space of complex \moire materials.

\end{abstract}

\maketitle

A small mismatch between the lattice vectors of adjacent two-dimensional (2D) materials generates a \moire pattern: an emergent superlattice that modulates the interlayer atomic registry on length scales much larger than the atomic lattice constant.
The twist angle degree of freedom enables continuous tuning of \moire potentials and hence electronic structures, establishing \moire materials as a designer platform for novel quantum phenomena, including correlated insulating states~\cite{Cao2018Correlated}, superconductivity~\cite{Cao2018Unconventional}, and integer and fractional quantum anomalous Hall effects~\cite{Serlin2020Intrinsic,Park2023Observation}.
Most prior studies have focused on systems with a single \moire pattern, such as twisted bilayer graphene, where correlated states emerge near discrete ``magic" twist angles that yield flat bands.
Multi-\moire systems, which host two or more distinct \moire patterns, provide additional twist angle degrees of freedom and enable the exploration of magic conditions in higher-dimensional twist-angle spaces.

\begin{figure*}[p]
    \centering
    \includegraphics[width=6.48in]{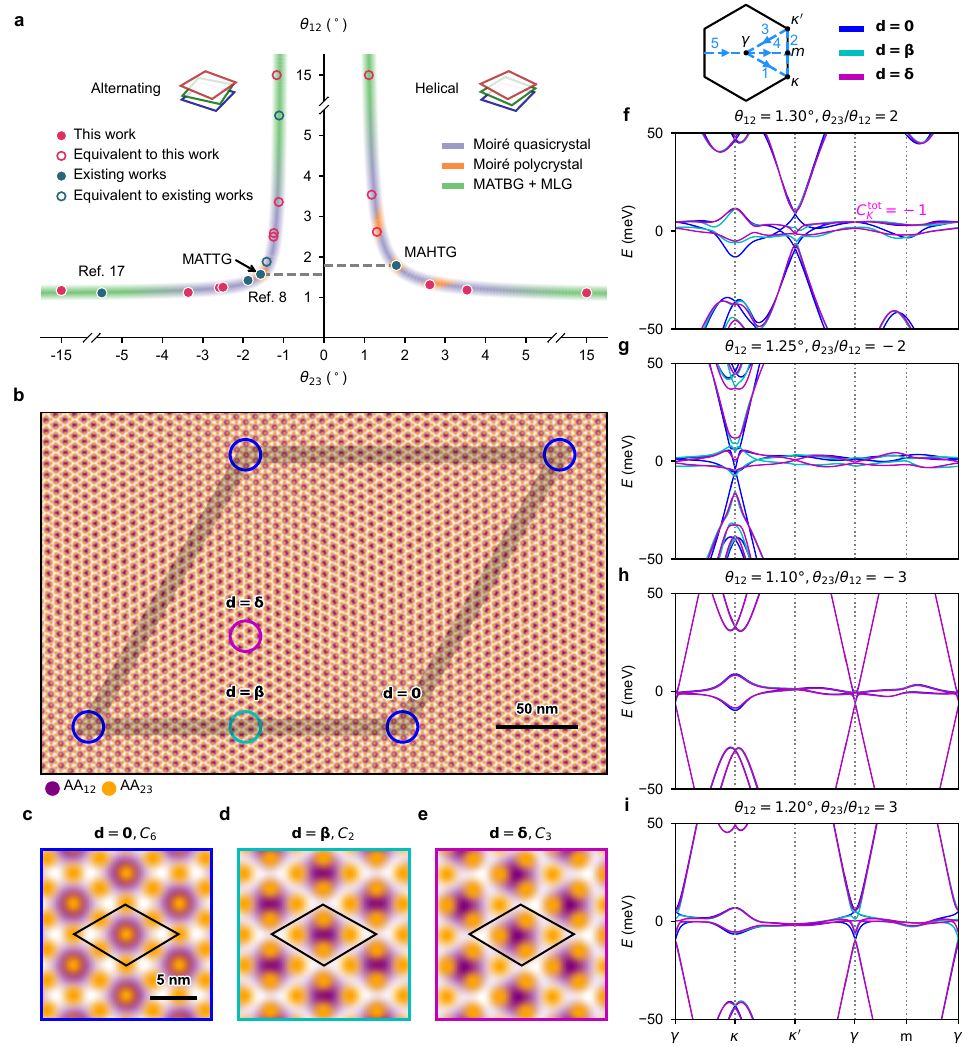}
    \caption{\figtitle{Magic continuum in TTG.}
    \panel{a}, Structural parameter space of TTG. 
    Solid circles and their hollow equivalents denote twist-angle combinations where strong correlations were observed in previous works (dark green) and in the present study (red). 
    They lie on two branches of a magic continuum in the parameter space.
    Based on the degree of lattice relaxation, TTG can be categorized into \moire quasicrystals (purple-shaded regions) and \moire polycrystals (orange-shaded regions).
    For large values of $|\theta_{12}|$ $\left(|\theta_{23}|\right)$ along the magic continuum, one of the layers becomes electronically decoupled and the other two layers form magic-angle twisted bilayer graphene where $|\theta_{23}|\,\left(|\theta_{12}|\right)\sim1.1^\circ$ (green-shaded regions).
    MATBG, magic-angle twisted bilayer graphene; MLG, monolayer graphene; MATTG, magic-angle (alternating and mirror-symmetric) twisted trilayer graphene; MAHTG, magic-angle helical trilayer graphene.
    \panel{b}, Unrelaxed structure of TTG with $(\theta_{12}, \theta_{23})=(2\degree, -4\degree)$.
    Purple and orange dots represent AA stacking of the top and bottom pairs of adjacent layers, respectively.
    Together they form a \supermoire pattern with a unit cell outlined by a grey diamond.
    Blue, cyan, and magenta circles mark high-symmetry points of the \supermoire unit cell, where the shifts between two \moire AA stacking sites are $\mathbf{0}$, $\boldsymbol{\beta}$, and $\boldsymbol{\delta}$, respectively.
    \panel{c-e}, Zoomed-in views of the \moire patterns at the three points marked in \panel{b}.
    Black diamonds outline the \moire unit cells under the periodic approximation.
    \panel{f-i}, Non-interacting band structures for different configurations along the magic continuum: $\theta_{12}:\theta_{23}=1:2$ (\panel{f}), $1:-2$ (\panel{g}), $1:-3$ (\panel{h}), and $1:3$ (\panel{i}).
    These calculations are performed with zero interlayer electric potential difference.
    Local \moire band structures at the various high-symmetry points within the \supermoire unit cell are shown; each is coloured to match its corresponding panel in \panel{c-e}.
    }
    \label{fig:1}
\end{figure*}

Here, we focus on one of the simplest multi-\moire systems---twisted trilayer graphene (TTG), in which three graphene monolayers are stacked with two independent twist angles: $\theta_{12}$ between layers 1 and 2, and $\theta_{23}$ between layers 2 and 3 (\figlbl~\ref{fig:1}\panel{a}).
This trilayer stacking yields either a helical or alternating twist configuration, depending on whether $\theta_{12}$ and $\theta_{23}$ have the same or opposite signs, respectively.
Except in the special cases of mirror-symmetric TTG ($\theta_{12}=-\theta_{23}$) and twisted monolayer-bilayer graphene ($\theta_{12}=0$ or $\theta_{23}=0$), TTG generally hosts two incommensurate \moire patterns, resulting in a rich and tunable multi-\moire landscape.
Previous studies reported contrasting correlated phenomena in several specific configurations of TTG, including superconductivity near $(\theta_{12}, \theta_{23})\approx(1.6\degree, -1.6\degree)$ or $(1.4\degree, -1.9\degree)$ \cite{Park2021Tunable,Hao2021Electric,Uri2023Superconductivity}, and an anomalous Hall effect (AHE) near $(\theta_{12}, \theta_{23})\approx(1.8\degree, 1.8\degree)$~\cite{Xia2025Topological} (these works are denoted as dark green circles in \figlbl~\ref{fig:1}\panel{a}).
However, the underlying principles that govern the ground states of different twist angle configurations, and the rules that dictate where correlated phenomena arise in the two-angle space, remain unclear.

\begin{figure*}
    \centering
    \includegraphics[width=7in]{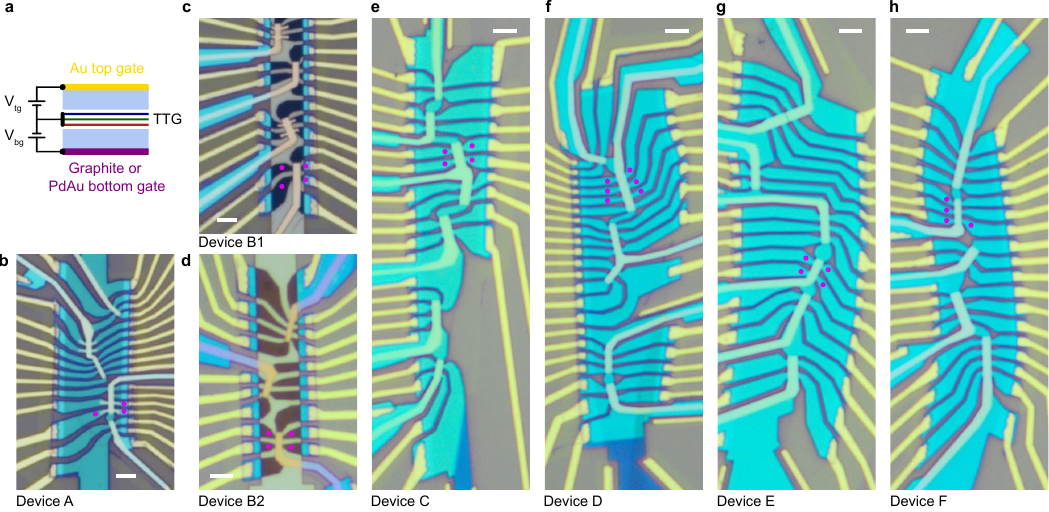}
    \caption{\figtitle{Circuit diagram and optical micrographs of TTG devices.}
    \panel{a}, Circuit diagram of TTG surrounded by two hBN dielectric layers and top and bottom gate electrodes (Au top gate and graphite or PdAu bottom gate) kept at electric potentials $V_\text{tg}$, $V_\text{bg}$ relative to TTG.
    \panel{b}, Device A.
    \panel{c}, Device B1.
    \panel{d}, Device B2.
    \panel{e}, Device C.
    \panel{f}, Device D.
    \panel{g}, Device E.
    \panel{h}, Device F.
    $R_{xx}$ and $R_{yx}$ contacts used in this study are indicated by magenta dots for all devices.
    All scale bars are $\SI{3}{\mathrm{\mu}m}$.
    \label{fig:devices}
    }
\end{figure*}

In this work, we observed correlated ground states in seven multi-\moire TTG devices with new twist angle configurations, denoted as red circles in \figlbl~\ref{fig:1}\panel{a} (see also Methods~\ref{ssec:fab}, Methods~\ref{ssec:transport}, and \extlbl~\ref{fig:devices}).
Remarkably, all samples with twist angles near two continuous lines in the parameter space exhibit either superconductivity or AHE.
This fact, together with the physical expectation that the electronic structure varies smoothly with twist angles, strongly suggests that correlated phases in TTG are not restricted to isolated magic points but persist along an extended manifold, which we refer to as ``magic continuum".

\section*{Electronic structure of multi-\moire TTG}

Due to the lack of \moire commensuration, a global \moire band structure does not exist for multi-\moire TTG.
Nonetheless, its electronic properties can be effectively modelled as spatially varying local \moire bands by approximating the structure as locally periodic on the \moire length scale.
This is implemented by approximating the two twist angles as a fraction, $(\theta_{12}, \theta_{23})\approx(p\theta_0, q\theta_0)$ with $p$ and $q$ being coprime integers, and applying a small structural distortion to render the two bilayer \moire wave vectors commensurate~\cite{Popov2023Magic,Yang2024Multi}.
Within this framework, a generalized Bistritzer–MacDonald continuum model~\cite{Bistritzer2011Moire} can be applied locally within each \moire unit cell of size $l_\text{m}\approx a_0/\theta_0$, where $a_0=\SI{0.246}{nm}$ is the graphene lattice constant.
The resulting local \moire bands are parametrized by the relative lateral displacement $\mathbf{d}$ between the two \moire lattices, which varies smoothly and periodically over the \supermoire (\moire of \moire) scale, $l_\mathrm{sm}\approx 2a_0/\left|(p+q)pq\theta_0^2\right|$.
Figure~\ref{fig:1}\panel{b} illustrates the unrelaxed structure of multi-\moire TTG for $(\theta_{12}, \theta_{23})=(2\degree,-4\degree)$, with the \supermoire unit cell outlined in gray.
Figures~\ref{fig:1}\panel{c-e} show the approximated \moire unit cell at three high-symmetry points within the \supermoire unit cell (indicated by coloured circles in \figlbl~\ref{fig:1}\panel{b}), with $\mathbf{d}$ labelled as $\mathbf{0}$, $\boldsymbol{\beta}$, and $\boldsymbol{\delta}$, respectively.
These distinct \moire patterns exhibit different symmetries depending on $\mathbf{d}$~\cite{Hao2024Robust}, and lead to a \supermoire-modulated local electronic structure.
This is illustrated in \figlbls~\ref{fig:1}\panel{f-i}, which present calculated non-interacting \moire band structures at the aforementioned $\mathbf{d}$ (indicated by different colours), for simple $(p, q)$ of $(1, 2)$ (\panel{f}), $(1, -2)$ (\panel{g}), $(1, -3)$ (\panel{h}), and $(1, 3)$ (\panel{i}) (Methods~\ref{ssec:calc}).

By maximizing the peak in the density of states averaged over different $\mathbf{d}$ values, one can determine a magic angle for each $(p, q)$ combination, presenting target heterostructures likely to host correlated phases.
Within the $(\theta_{12}, \theta_{23})$ parameter space, these magic angles for various $(p, q)$ ratios lie along two branches of a magic continuum that extend through the alternating and helical twist quadrants, as shown in \figlbl~\ref{fig:1}\panel{a}~\cite{Zhu2020Twisted,Foo2023Extended,Popov2023Magic,Yang2024Multi}.
Exchanging $\theta_{12}$ and $\theta_{23}$ ($-\theta_{23}$) yields equivalent helical (alternating) structures.
Therefore, the helical and alternating branches of the magic continuum are mirror symmetric about the lines $\theta_{12}=\theta_{23}$ and $\theta_{12}=-\theta_{23}$, respectively.
In the following, we assume $\left|\theta_{23}/\theta_{12}\right|\geq1$ without loss of generality.
Notably, the two branches are not mirror images of each other about the $\theta_{23}=0$ axis.
For instance, the magic angle for $\theta_{12}=\theta_{23}$ is around $1.8\degree$, whereas for $\theta_{12}=-\theta_{23}$, it is around $1.6\degree$.

\begin{figure*}
    \centering
    \includegraphics[width=14cm]{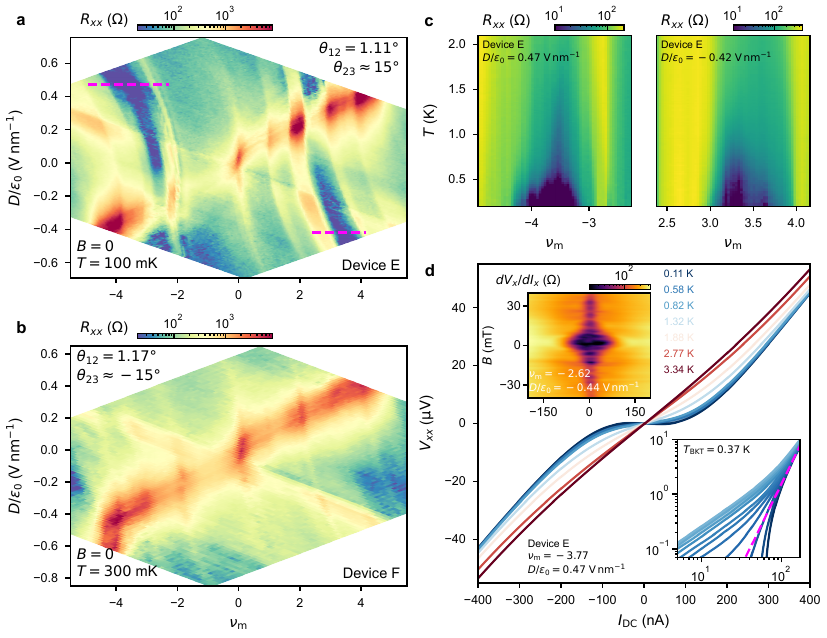}
    \caption{\figtitle{Correlated states and superconductivity in TTG with one graphene layer electronically decoupled.}
    \panel{a}, $R_{xx}$ versus $\nu_\mathrm{m}$ and $D$ measured at $B=0$ and $T=\SI{300}{mK}$ for Device E with $\theta_{12}=1.11\degree, \theta_{23}\approx 15\degree$ (three layers of graphene helically twisted), showing resistance peaks at charge neutrality ($\nu_\mathrm{m}=0$), at the full filling of the first \moire bands ($\nu_\mathrm{m}=\pm 4$), and at correlated states ($\nu_\mathrm{m}=-2, 1, 2, 3$).
    The blue regions signal superconductivity.
    \panel{b}, Same as \panel{a}, measured at $T=\SI{300}{mK}$ for Device F with $\theta_{12}=1.17\degree, \theta_{23}\approx -15\degree$ (three layers of graphene alternating twisted), showing correlated states with peaked $R_{xx}$ at $\nu_\mathrm{m}=\pm 2$.
    The lack of signatures for superconductivity is possibly due to the higher $T$ or sample not having the optimal twist angle.
    \panel{c}, $R_{xx}$ versus $\nu_\mathrm{m}$ and $T$ measured for Device E along the magenta lines marked in \panel{a}, demonstrating superconducting transitions on both electron- and hole-doped sides.
    Data is taken at $D/\epsilon_0=\SI{0.47}{V\,nm^{-1}}$ (left) and $-\SI{0.42}{V\,nm^{-1}}$ (right).
    \panel{d}, $V_{xx}$ versus $I_\mathrm{DC}$ curves at various $T$, measured in Device E.
    Data is taken at $\nu_\mathrm{m}=-3.77$, $D/\epsilon_0=\SI{0.47}{V\,nm^{-1}}$.
    Bottom-right insets plot the same data in the log-log scale, sampled at finer temperature increments, where $T_\mathrm{BKT}=\SI{0.37}{K}$ can be extracted.
    Top-left insets show differential resistance $dV_{x}/dI_{x}$ versus $I_\mathrm{DC}$ and small perpendicular magnetic field $B$, demonstrating Josephson interference patterns, which are evidence for robust superconductivity.
    \label{fig:MATBG+MLG}
    }
\end{figure*}

As either $|\theta_{12}|$ or $|\theta_{23}|$ increases, both branches of the magic continuum asymptotically approach the limit of magic-angle twisted bilayer graphene stacked on an electronically decoupled monolayer~\cite{Hoke2024Uncovering} (green-shaded regions in \figlbl~\ref{fig:1}\panel{a}), in which the local low-energy electronic properties have a vanishing dependence on $\mathbf{d}$ (note that the decoupled layer is still expected to affect the bilayer via screening).
This reduced $\mathbf{d}$-dependence is evident when comparing the local \moire band structures for $\left|\theta_{23}/\theta_{12}\right|=3$ (\figlbls~\ref{fig:1}\panel{h,i}) with those for $\left|\theta_{23}/\theta_{12}\right|=2$ (\figlbls~\ref{fig:1}\panel{f,g}).
Below, we focus on the four structures with $(\left|\theta_{23}/\theta_{12}\right|\approx 2, 3)$, in which the \supermoire modulation of the local electronic properties plays a more important role.
Results from two additional samples with $|\theta_{23}|\approx 15\degree$ are summarized in \extlbl~\ref{fig:MATBG+MLG}, serving as examples in which the third layer is electronically decoupled (see Methods~\ref{ssec:twist} for a description of the twist angle determination).

\section*{Correlations and AHE in TTG \moire polycrystals}

Lattice relaxation in TTG depends on the specific angular configuration and can strongly reshape the \supermoire landscape.
When $\theta_{23}/\theta_{12}$ is close to $\pm 1$ or $2$, TTG is predicted to relax to \moire-periodic domains separated by sharp domain walls on the \supermoire scale, resembling the structure of a polycrystal \cite{Yang2024Multi} (orange-shaded regions in \figlbl~\ref{fig:1}\panel{a}).
While \moire polycrystals with $\left|\theta_{23}/\theta_{12}\right|\approx1$ have been studied experimentally~\cite{Park2021Tunable,Hao2021Electric,Xia2025Topological,Hoke2024Imaging}, here we report the first investigation of a \moire polycrystal with $\theta_{23}/\theta_{12}\approx2$.
\begin{figure*}[t]
    \centering
    \includegraphics[width=7in]{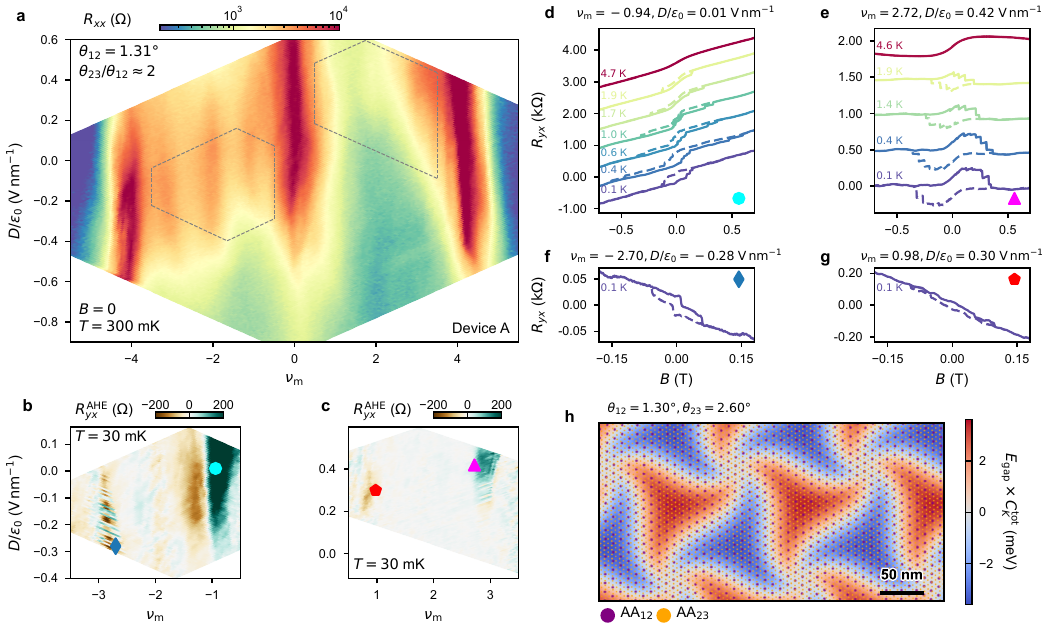}
    \caption{\figtitle{Correlated states and AHE in polycrystalline TTG.}
    \panel{a}, $R_{xx}$ versus $\nu_\mathrm{m}$ and $D$ measured at $B=0$ and $T=\SI{300}{mK}$ for a \moire polycrystal TTG device with $\theta_{23}/\theta_{12}\approx 2$ (Device A), showing resistance peaks at charge neutrality ($\nu_\mathrm{m}=0$), at the full filling of the first \moire bands ($\nu_\mathrm{m}=\pm 4$), and at correlated states ($\nu=-1,-2,-3$).
    \panel{b,c}, $R_{yx}^\mathrm{AHE}$ (see Methods~\ref{ssec:transport}) measured at $T=\SI{30}{mK}$ in hole- (\panel{b}) and electron-doped (\panel{c}) regions outlined by grey dashed polygons in \panel{a}.
    Regions of large $R_{yx}^\mathrm{AHE}$ at $\nu_\mathrm{m}=\pm 1,\pm 3$ indicate AHE.
    \panel{d}, Field-antisymmetrized $R_{yx}$ taken at $\nu_\mathrm{m}=-0.94$ and $D/\epsilon_0=\SI{0.01}{V\,nm^{-1}}$ (cyan circle in \panel{b}) while sweeping $B$ up (solid) and down (dashed) at different temperatures as indicated, demonstrating an AHE.
    Curves at different temperatures are shifted by $\SI{0.6}{k\Omega}$ for clarity.
    \panel{e}, Same as \panel{d}, taken at $\nu_\mathrm{m}=2.72$, $D/\epsilon_0=\SI{0.42}{V\,nm^{-1}}$ (pink triangle in \panel{c}).
    Curves at different temperatures are shifted by $\SI{0.48}{k\Omega}$ for clarity.
    \panel{f}, Same as \panel{d}, taken at $\nu_\mathrm{m}=-2.70$, $D/\epsilon_0=-\SI{0.28}{V\,nm^{-1}}$, $T=\SI{0.1}{K}$ (blue diamond in \panel{b}).
    \panel{g}, Same as \panel{d}, taken at $\nu_\mathrm{m}=0.98$, $D/\epsilon_0=\SI{0.30}{V\,nm^{-1}}$, $T=\SI{0.1}{K}$ (red pentagon in \panel{c}).
    \panel{h}, Calculated \supermoire structure of TTG with $\theta_{12}=1.30\degree, \theta_{23}=2.60\degree$, including lattice relaxation.
    As in \figlbl~\ref{fig:1}\panel{b}, purple and orange dots represent AA stacking of the top and bottom pairs of adjacent layers, respectively.
    Lattice relaxation leads to the formation of \moire periodic domains with $\mathbf{d}=\pm\boldsymbol{\delta}$ that host topological flat bands.
    Background colour represents $E_\mathrm{gap} \times C_K^\mathrm{tot}$ calculated for the local \moire lattices, where $C_K^\mathrm{tot}$ is the total Chern number per spin of the pair of flat bands in valley $K$, and $E_\mathrm{gap}\geq 0$ is the minimum direct band gap at $\nu_\mathrm{m}=\pm 4$.
    Red (blue) hues represent $C_K^\mathrm{tot}=1$ ($C_K^\mathrm{tot}=-1$), whereas gray indicates gapless domain walls.
    }
    \label{fig:2}
\end{figure*}

Figure~\ref{fig:2}\panel{a} shows the longitudinal resistance $R_{xx}$ measured at $B=0$ and $T=\SI{300}{mK}$ as a function of vertical displacement field, $D$, and number of electrons per \moire unit cell, $\nu_\mathrm{m}$.
Here, $B$ and $T$ denote the perpendicular magnetic field and temperature, respectively.
The \moire unit cell was defined in the previous section using the local periodic approximation.
In devices where either $p=1$ or $q=1$, this unit cell coincides with that of the small-angle bilayer \moire lattice (see, for example, \figlbls~\ref{fig:1}\panel{c-e}).
The large $R_{xx}$ peaks at $\nu_\mathrm{m}=\pm 4$ originate from the reduced density of states when the flat \moire bands are empty or full.
Additional $R_{xx}$ peaks appear at $\nu_\mathrm{m}=-1, -2$ and $-3$, indicating spin and valley degeneracy lifting, driven by electronic correlations.
To search for topologically non-trivial phases, we first measured the anomalous Hall response, $R_{yx}^\mathrm{AHE}$, at small $B$, as illustrated in \figlbls~\ref{fig:2}\panel{b, c} (see Methods~\ref{ssec:transport} for the measurement details).
Pronounced values of $\left|R_{yx}^\mathrm{AHE}\right|$ near $\nu_\mathrm{m}=\pm 1$ and $\pm 3$ signal AHE, corroborated by field sweep measurements shown in \figlbls~\ref{fig:2}\panel{d-g}.
Hysteresis of the field-antisymmetrized Hall resistance, $R_{yx}$, around zero magnetic field reveals a ferromagnetic ground state (see Methods~\ref{ssec:transport} for a description of the antisymmetrization procedure and \extlbl~\ref{fig:AHE_raw} for the raw data).
The temperature dependence of $R_{yx}$ curves near $\nu_\mathrm{m}=-1$ and $3$ indicates a Curie temperature between $\SI{1.9}{K}$ and $\SI{4.6}{K}$.
Similar to other carbon-based materials, due to the weak spin-orbit coupling, the ferromagnetism observed here likely has an orbital origin~\cite{Tschirhart2021Imaging,Sharpe2021Evidence}, a manifestation of non-zero Berry curvature.

Zero-field topological bands have not been observed in single-\moire 2D systems with global $xy$-inversion ($C_{2z}$) symmetry.
In contrast, despite the global $C_{2z}$ symmetry of our system, the multi-\moire structure combined with the aforementioned lattice relaxation leads to mesoscopic domains with uniform $\mathbf{d}=\pm\boldsymbol{\delta}$, which locally break $C_{2z}$ symmetry (see \figlbl~\ref{fig:2}\panel{h} for the relaxed structure calculated using the method of ref.~\cite{Carr2018Relaxation,Cazeaux2019Energy,Zhu2020Modeling}).
This behaviour is analogous to helical trilayer graphene with $\theta_{12}\approx\theta_{23}$~\cite{Xia2025Topological}.
Indeed, our band structure calculations produce nearly flat \moire bands with non-zero valley Chern numbers within each domain (\figlbl~\ref{fig:1}\panel{f}, magenta).
The observed AHE around odd integer fillings indicates spontaneous time-reversal symmetry breaking, likely originating from interaction-driven valley polarization or imbalance.
Since neighbouring \supermoire domains are related by a $C_{2z}$ operation, their local \moire flat bands possess opposite valley Chern numbers.
When globally polarizing to the same valley, this results in a ``Chern mosaic" (see \figlbl~\ref{fig:2}\panel{h})~\cite{Grover2022Chern,Devakul2023Magic-angle,Guerci2023Chern}.
These Chern domains are separated by topologically protected gapless domain walls, consistent with the absence of quantized AHE in the experiment.
We note that while the size and shape of the domains are extremely susceptible to strain and twist angle disorder~\cite{Hoke2024Imaging}, the band topology is expected to be robust.

\begin{figure*}
    \centering
    \includegraphics[width=7in]{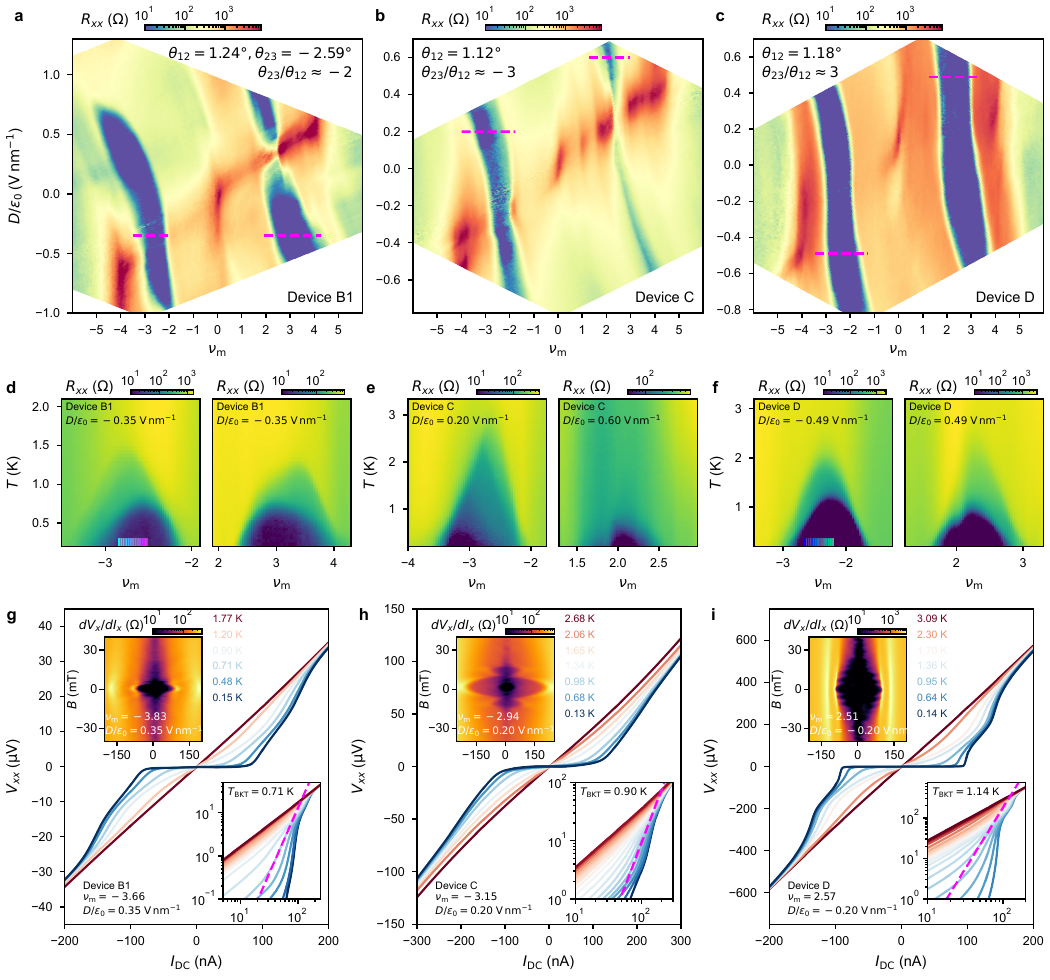}
    \caption{\figtitle{Correlated states and robust superconductivity in quasicrystalline TTG.}
    \panel{a-c}, $R_{xx}$ versus $\nu_\mathrm{m}$ and $D$ measured for devices with $\theta_{12}:\theta_{23}\approx 1:-2$ (Device B1, \panel{a}), $\theta_{12}:\theta_{23}\approx 1:-3$ (Device C, \panel{b}), and $\theta_{12}:\theta_{23}\approx 1:3$ (Device D, \panel{c}), showing resistance peaks at charge neutrality ($\nu_\mathrm{m}=0$), at the full filling of the first \moire bands ($\nu_\mathrm{m}=\pm 4$), and various correlated states at integer and fractional fillings.
    Blue regions indicate superconductivity.
    All data was taken at $B=0$ and $T\approx\SI{100}{mK}$.
    \panel{d-f}, $R_{xx}$ versus $\nu_\mathrm{m}$ and $T$ measured for Device B1 (\panel{d}), Device C (\panel{e}), and Device D (\panel{f}), along the magenta lines marked in \panel{a-c} ($D$ values indicated in \panel{d-f}), demonstrating superconducting transitions for all three quasicrystalline TTG devices on both electron- and hole-doped sides with critical temperatures on the order of $\SI{1}{K}$.
    \panel{g-i}, $V_{xx}$ versus $I_\mathrm{DC}$ curves at various $T$, for the corresponding devices in \panel{a-c}.
    $\nu_\mathrm{m}$ and $D$ values are indicated.
    Bottom-right insets show the same data in log-log scale, sampled at finer temperature increments.
    By fitting to $V_{xx}\propto I_\mathrm{DC}^3$ (dashed magenta lines), we extract $T_\mathrm{BKT}=\SI{0.71}{K}, \SI{0.90}{K}$, and $\SI{1.14}{K}$ for the three devices, respectively.
    Top-left insets show differential resistance $dV_{x}/dI_{x}$ versus $I_\mathrm{DC}$ and small $B$, demonstrating Josephson interference patterns---evidence for phase coherent transport.
    }
    \label{fig:3}
\end{figure*}

\section*{Correlations and superconductivity in TTG \moire quasicrystals}

When $\theta_{23}/\theta_{12}$ is away from $\pm 1$ and $2$, lattice relaxation is expected to have minimal impact on the \supermoire landscape, i.e., $\mathbf{d}$ evolves smoothly.
The resulting structure has been referred to as a ``\moire quasicrystal"~\cite{Uri2023Superconductivity} due to the incommensuration between the two \moire lattices.

\begin{figure*}
    \centering
    \includegraphics[width=14cm]{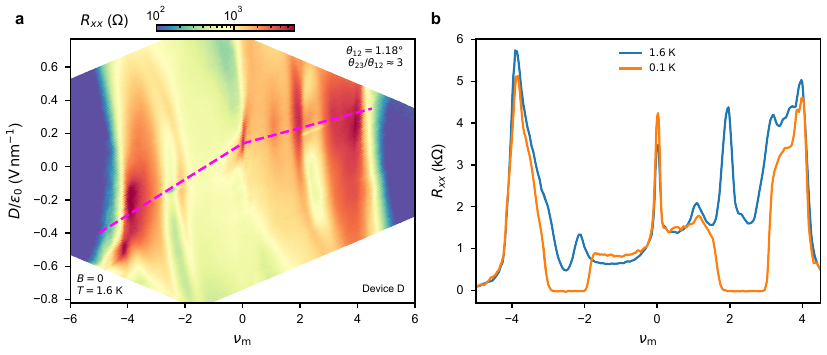}
    \caption{\figtitle{Correlated states in quasicrystalline TTG with $\theta_{23}/\theta_{12}\approx 3$.}
    \panel{a}, $R_{xx}$ versus $\nu_\mathrm{m}$ and $D$ measured for Device D at $B=0$ and $T=\SI{1.6}{K}$.
    Spin/valley degeneracy breaking states driven by electronic correlations manifest as $R_{xx}$ peaks at $\nu_\mathrm{m}=-2, 1, 2$ and $3$.
    \panel{b}, Line cuts of $R_{xx}$ along the magenta line shown in \panel{a}, plotted versus $\nu_\mathrm{m}$.
    Blue and orange curves are taken at $T=\SI{1.6}{K}$ and $T=\SI{0.1}{K}$, respectively.
    At lower temperature, most correlated states evolve into superconductivity.
    \label{fig:ML11_highT}
    }
\end{figure*}

Here, we present a systematic study of quasicrystalline TTG along both helical and alternating branches of the magic continuum, showing results from devices close to simple angle ratios $\theta_{23}/\theta_{12}=-2, -3$ and $3$.
Figures~\ref{fig:3}\panel{a-c} show $R_{xx}$ versus $\nu_\mathrm{m}$ and $D$ measured at $B=0$ and $T=\SI{100}{mK}$.
Remarkably, all three structures demonstrate superconductivity, indicated by the deep blue areas.
In addition to superconductivity, a series of correlated states at integer and fractional fillings manifests as peaks in $R_{xx}$, as summarized in \exttbllbl~\ref{table:CS} (see \extlbl~\ref{fig:Landau_fan} and Methods~\ref{ssec:Landau_fan} for correlated states under finite magnetic field).
Local dispersive \moire bands coexist with local flat bands in quasicrystalline TTG (\figlbls~\ref{fig:1}\panel{g-i}), allowing the former to shunt high-resistance correlated states supported by the latter.
As a result, high-resistance states are prominent only along a diagonal line, where the Fermi level aligns with a low density of states in the local dispersive bands.
In Device D (shown in \figlbl~\ref{fig:3}\panel{c}), correlated high-resistance states are weak or absent, which may result from competition with superconductivity.
Extended~Data~Figure~\ref{fig:ML11_highT} shows the same measurement performed at elevated temperature, where the correlated states become clearly resolved.
Identifying the spin, valley, and sublattice order of different correlated states remains an open question for future theoretical and experimental studies.
The many-body ground states are likely spatially modulated within the \supermoire unit cell due to the $\mathbf{d}$-dependent \moire band structure, making local probes such as scanning tunnelling microscopy well-suited for this task~\cite{Nuckolls2023Quantum,Kim2023Imaging}.

To further explore the superconducting properties of quasicrystalline TTG, we measured $R_{xx}$ versus $\nu_\mathrm{m}$ and $T$ at constant $D$, showing characteristic superconducting domes with transition temperatures $T_\mathrm{c}$ around $\SI{1}{K}$ (\figlbls~\ref{fig:3}\panel{d-f}).
Figures~\ref{fig:3}\panel{g-i} present the DC voltage-current ($V_{xx}$–$I_\mathrm{DC}$) characteristics at selected values of $\nu_\mathrm{m}$ and $D$ as a function of temperature.
As $T$ decreases, the $V_{xx}$–$I_\mathrm{DC}$ curves evolve from a linear (Ohmic) regime to a non-linear profile, characteristic of superconductivity.
By fitting to $V_{xx}\propto I_\mathrm{DC}^3$, we extract the Berezinskii–Kosterlitz–Thouless transition temperature, $T_\mathrm{BKT}$, as shown in the bottom-right insets of \figlbls~\ref{fig:3}\panel{g-i}.
Furthermore, $V_{xx}$–$I_\mathrm{DC}$ measurements under a small perpendicular magnetic field reveal Fraunhofer-like interference patterns (\figlbls~\ref{fig:3}\panel{g-i}, top-left insets), providing compelling evidence for the presence of phase-coherent superconductivity.

\section*{Signatures for \Supermoire-modulated superconductivity}

\begin{figure}
    \centering
    \includegraphics[width=3.38in]{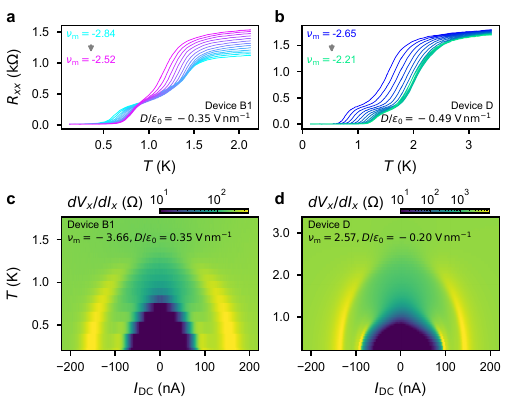}
    \caption{\figtitle{Signatures for \supermoire-modulated superconductivity in quasicrystalline TTG.}
    \panel{a-b}, $R_{xx}$ versus $T$ measured in Device B1 (\panel{a}) and Device D (\panel{b}), both demonstrating a two-step behaviour of the superconducting transition.
    Curve colours in \panel{a} (\panel{b}) match the tick colours at the bottom of \figlbl~\ref{fig:3}\panel{d} left (\figlbl~\ref{fig:3}\panel{f} left), indicating $\nu_\mathrm{m}$ and $D$ where the data was taken.
    \panel{c}, $dV_{x}/dI_{x}$ versus $I_\mathrm{DC}$ and $T$ measured in Device B1 ($\nu_\mathrm{m}$ and $D$ values are indicated).
    \panel{d}, Same as \panel{c}, measured in Device D.
    For both devices, two superconducting coherence peaks appear below the two superconducting transition temperatures, respectively.
    Non-linear resistance with coherence peaks indicates the existence of Cooper pairs in the intermediate temperature regime between the two transitions.
    }
    \label{fig:4}
\end{figure}

\begin{figure*}
    \centering
    \includegraphics[width=7in]{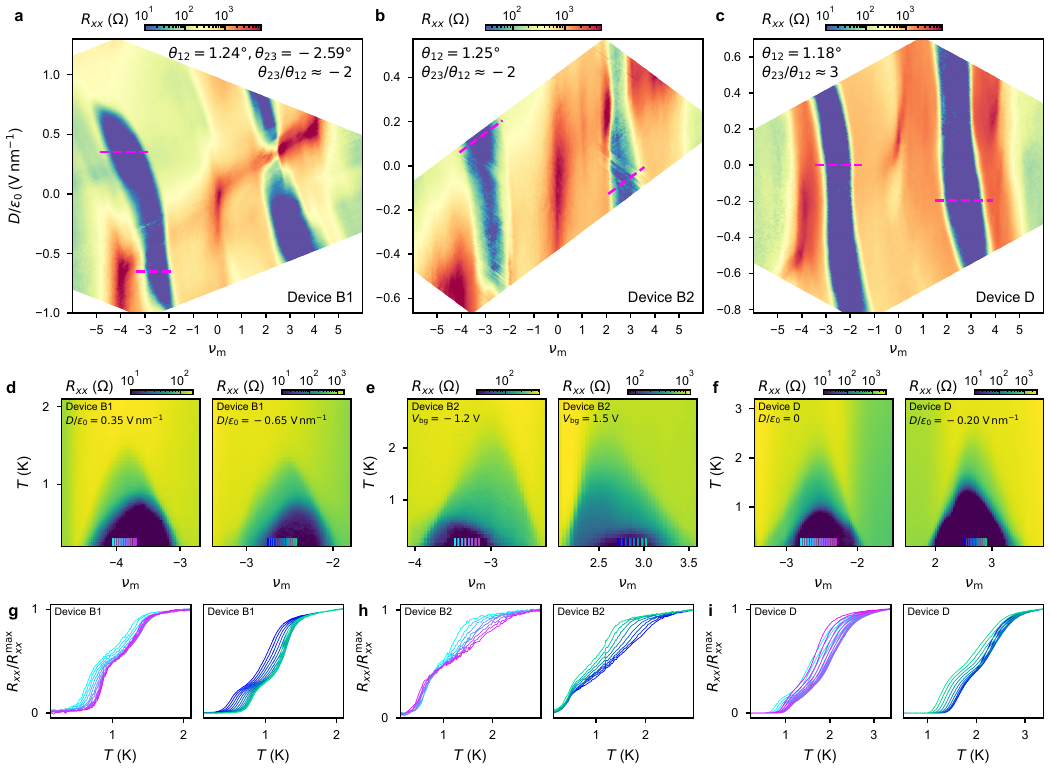}
    \caption{\figtitle{Ubiquitous two-dome structure of superconductivity in TTG devices with $\theta_{23}/\theta_{12}\approx -2$ and $3$.}
    \panel{a-c}, $R_{xx}$ versus $\nu_\mathrm{m}$ and $D$ measured at $B=0$ and $T\approx\SI{100}{mK}$ for Device B1 (\panel{a}), Device B2 (\panel{b}), and Device D (\panel{c}).
    Data in \panel{a,c} is reproduced from \figlbls~\ref{fig:3}\panel{a,c}.
    \panel{d-f}, $R_{xx}$ versus $\nu_\mathrm{m}$ and $T$ measured for Device B1, Device B2, and Device D along the magenta lines marked in \panel{a-c}.
    All exhibit two-dome structures of the superconducting transition.
    \panel{g-i}, $R_{xx}$ versus $T$ curves measured for Device B1 (\panel{g}), Device B2 (\panel{h}), and Device D (\panel{i}), all illustrating two-step behaviour for the superconducting transition.
    The colours of the curves match the colours of the ticks at the bottom of \panel{d-f}, indicating the $\nu_\mathrm{m}$ and $D$ where the data is taken.
    \label{fig:2dome}
    }
\end{figure*}

In Devices B1 and D, the superconducting regions exhibit a two-dome structure as a function of $\nu_\mathrm{m}$ and $T$, shown in \figlbls~\ref{fig:3}\panel{d,f}.
This behaviour is further substantiated by the $R_{xx}(T)$ traces in \figlbls~\ref{fig:4}\panel{a,b}, which exhibit two distinct transition temperatures where the resistance sharply drops upon cooling.
Notably, this two-dome behaviour is a ubiquitous feature for devices with $\theta_{23}/\theta_{12}\approx -2$ and $3$ (see \extlbl~\ref{fig:2dome} for more results from Devices B1, B2, and D), strongly indicating that it is intrinsic to the system rather than arising from extrinsic factors such as disorder.
Between the two transitions, $R_{xx}(T)$ traces can be well described by the superconducting proximity effect model (Methods~\ref{ssec:SM_SC} and \extlbl~\ref{fig:RTfit})~\cite{Abraham1982Resistive}, suggesting the existence of superconductor-normal metal-superconductor junctions.
Furthermore, we measured differential resistance $dV_{x}/dI_{x}$ as a function of $I_\mathrm{DC}$ and $T$ (shown in \figlbls~\ref{fig:4}\panel{c,d} and \extlbl~\ref{fig:IVcut}) and magnetoresistance $R_{xx}(B)$ as a function of $T$ (shown in \extlbl~\ref{fig:MR}), both of which indicate that part of the sample remains superconducting between the two transition temperatures (see detailed discussion in Methods~\ref{ssec:SM_SC}).

\begin{figure*}
    \centering
    \includegraphics[width=3.38in]{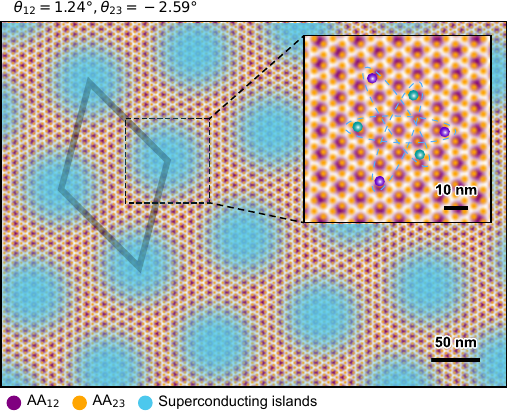}
    \caption{\figtitle{Schematic of \supermoire-modulated superconductivity in quasicrystalline TTG.}
    \Moire AA sites and \supermoire unit cell are labelled as in \figlbl~\ref{fig:1}\panel{b}.
    We use the twist angle values experimentally extracted for Device B1. 
    Due to the \supermoire-modulated local electronic structure, certain regions in the \supermoire unit cell---such as $\mathbf{d}=\mathbf{0}$ shown here---can host superconductivity without global phase coherence between the two superconducting transitions.
    Inset shows zoomed-in illustration of the superconducting region, which is larger than or comparable to the size of Cooper pairs, given by the experimentally extracted Ginzburg–Landau coherence length.
    }
    \label{fig:SM_SC}
\end{figure*}

Similar transport behaviour was reported in 2D systems with patterned superconducting islands~\cite{Resnick1981Kosterlitz-Thouless,Abraham1982Resistive,Eley2012Approaching,Han2014Collapse,Bøttcher2018Superconducting}, where the higher transition signals the onset of superconductivity within each island, and the lower transition corresponds to the establishment of global phase coherence across islands.
A highly likely explanation of our results is that the superconductivity in these devices is spatially modulated on the \supermoire length scale, naturally originating from the $\mathbf{d}$-dependent local electronic structure (see schematic in \extlbl~\ref{fig:SM_SC}).
At the higher transition temperature, certain regions within the \supermoire unit cell---such as the area near $\mathbf{d}=\mathbf{0}$ in the schematic---become superconducting, resulting in a sharp drop in $R_{xx}$.
Upon further cooling, two scenarios may occur at the lower transition temperature: (1) Josephson coupling between isolated superconducting regions becomes sufficiently strong to establish global phase coherence; or (2) the superconducting regions expand and form a percolating network of intrinsic superconductors.
In both cases, these developments lead to a vanishing $R_{xx}$.
Further discussion of this physical picture is provided in Methods~\ref{ssec:SM_SC}.

\section*{Outlook}

Single \moire twisted graphene systems can be divided into two families: alternating twisted graphene~\cite{Cao2018Correlated,Cao2018Unconventional,Park2021Tunable,Hao2021Electric,Park2022Robust,Zhang2022Promotion,Burg2022Emergence} (including twisted bilayer graphene) and twisted M+N-layer graphene~\cite{Cao2020Tunable,Liu2020Tunable,Polshyn2020electrical,Waters2024Topological} (M, N, or both are larger than 1).
Although correlated states emerge in both families when the interlayer twist angles are appropriate, robust superconductivity has thus far been observed only in the former~\cite{Cao2018Unconventional,Park2021Tunable,Hao2021Electric,Park2022Robust,Zhang2022Promotion,Burg2022Emergence}, whereas the latter is typically characterized by Chern bands and usually exhibits the AHE~\cite{Polshyn2020electrical,Waters2024Topological}.
While non-trivial band topology is enabled by the absence of $C_{2z}$ symmetry, the underlying conditions for superconductivity in \moire graphene systems require more investigation.
Our systematic study of TTG indicates a phenomenological trend along the magic continuum: helically twisted \moire polycrystals, dominated by regions with $\mathbf{d} = \pm\boldsymbol{\delta}$, consistently host AHE, whereas robust superconductivity emerges in TTG \moire quasicrystals, where $\mathbf{d}$ varies smoothly across the \supermoire unit cell.
This observation suggests that superconductivity is suppressed in regions where $\mathbf{d} = \pm\boldsymbol{\delta}$, despite the presence of local flat \moire bands.
Thus, a high density of states alone is insufficient to drive superconductivity in graphene \moire systems, suggesting that additional factors---such as local symmetries---play an important role.

Furthermore, our results reveal the profound effects of the \supermoire.
It gives rise to AHE in globally-$C_{2z}$-symmetric \moire polycrystals and modulates superconductivity in \moire quasicrystals.
The latter, in particular, provides a new platform for studying the quantum breakdown of superconductivity~\cite{Sacépé2020Quantum}.
While previous studies have constructed proximity-coupled arrays of superconducting islands to probe such transitions, \moire quasicrystals offer new opportunities due to several unique properties: (1) superconducting regions are naturally defined by atomically smooth, spatially varying \moire potentials rather than by lithographic patterning, (2) the characteristic size of superconducting regions is comparable to the superconducting coherence length, (3) the superconductivity is likely unconventional, as in other superconducting graphene \moire systems~\cite{Cao2021Pauli-limit,Oh2021Evidence,Kim2022Evidence,Tanaka2025Superfluid,Banerjee2025Superfluid,Park2025Simultaneous}, and (4) regions with broken $C_{2z}$ symmetry may host non-trivial local band topology that alters Josephson coupling.
These features, combined with the exceptional tunability of 2D \moire materials, position \moire quasicrystals as a promising platform for exploring superconductor-metal transitions, anomalous metallic phases, and related emergent quantum phenomena.

Lastly, the concept of a magic continuum highlights that correlated phenomena are not confined to isolated ``magic points," but can persist across extended regions---such as ``magic hyperplanes"---in multi-dimensional parameter spaces spanned by continuous variables including multiple twist angles, strain, and pressure.
This perspective offers a versatile route to exploring and engineering novel quantum phases~\cite{Carr2018Pressure,Yankowitz2019Tuning,Morales2023Pressure-enhanced}.
With the advent of in-situ techniques capable of simultaneously tuning multiple degrees of freedom in 2D materials~\cite{Inbar2023The,Tang2024On-chip}, the landscape of accessible strongly correlated systems can be greatly expanded.

\clearpage
\onecolumngrid
\section*{Methods}

\subsection{Device fabrication}\label{ssec:fab}
The van der Waals heterostructures were assembled in two parts using a standard dry-transfer technique~\cite{Sun2025Optimized}.
First, a hexagonal boron nitride (hBN) flake (usually with a thickness of \SI{15}{\nano m} - \SI{30}{\nano m}) and a few-layer graphene strip were picked up by a poly(bisphenol A carbonate)/polydimethylsiloxane stamp. 
This bottom stack and poly(bisphenol A carbonate) film were released onto a $\SI{285}{nm}$ SiO$_2$/Si substrate. 
For devices with metallic bottom gates, only an hBN flake was picked up and released onto a pre-patterned metallic strip (\SI{15}{nm} Pd/Au alloy with a \SI{2}{nm} Cr or Ti adhesion layer). 
After dissolving the poly(bisphenol A carbonate) film in chloroform, the bottom stack was annealed at \SI{350}{\degree C} under vacuum for 12 hours to remove polymer residues. 
Then, tip cleaning was performed using the contact mode of an atomic force microscope to further clean the surface.

For assembling the twisted trilayer graphene, a monolayer graphene flake was cut into three pieces using an optical microscope with a fibre-coupled supercontinuum laser.
A second poly(bisphenol A carbonate)/polydimethylsiloxane stamp was used to pick up an hBN flake and the three graphene pieces subsequently. 
Before the second and third pieces of graphene were picked up, the stage was rotated to realize the desired interlayer twist angle. 
Each layer of graphene was picked up slowly and at room temperature to minimize any unintentional perturbation of the twist angles. 
After assembling the top stack, it was released onto the bottom stack at \SI{150}{\degree C} - \SI{170}{\degree C}.
This release step was done quickly to minimize the time spent at elevated temperatures.

The Hall bar was defined in a bubble-free region of the completed heterostructure, identified using atomic force microscopy. 
Patterns were defined using an Elionix ELS-HS50 electron-beam lithography system. 
A metallic top gate (\SI{25}{nm} - \SI{65}{nm} Au with a \SI{2}{nm} - \SI{5}{nm} Cr or Ti adhesion layer) was deposited using a Sharon thermal evaporator. 
The device was connected using one-dimensional contacts (\SI{63}{nm} - \SI{75}{nm} Au with a \SI{2}{nm} - \SI{5}{nm} Cr adhesion layer)~\cite{Wang2013One-dimensional}. 
Finally, the device was etched into a Hall bar geometry using reactive-ion etching.

\subsection{Electrical transport measurements}\label{ssec:transport}
Pre-characterization of the devices was carried out in a Janis helium-3 refrigerator with an $\SI{8}{T}$ perpendicular superconducting magnet and a base temperature of about $\SI{290}{mK}$.
A home-made $\SI{65}{cm}$ twisted-pair copper tape filter with $\sim\SI{20}{MHz}$ cut-off frequency \cite{Spietz2006Twisted} was thermally anchored at the helium-3 pot to prevent heating from Johnson noise and reduce the electron temperature.
DC voltages were applied to the top and bottom gates using Keithley 2400/2450 source-measure units.
The AC excitation of $\SI{1}{nA}-\SI{10}{nA}$ at $\SI{10}{Hz}-\SI{25}{Hz}$ was applied using Stanford Research Systems SR830 or SR860 lock-in amplifiers.
The corresponding AC currents and voltages were preamplified by DL-1211 current preamplifiers and DL-1201/SR560 voltage preamplifiers respectively, then measured by the lock-in amplifiers.
The temperature was measured using a calibrated CX-1010-CU-HT-0.1L thermometer.

A portion of the dilution refrigerator measurements were performed in a Bluefors LD250 using an Attocube two-way rotation probe. 
Two filters are installed and thermally anchored to the mixing chamber stage to reduce the electron temperature: a Quantum Machines Qfilter with a $\SI{65}{\kilo Hz}$ RC circuit and a $\SI{225}{\mega Hz}$ LC circuit, and a home-made twisted-pair copper tape filter similar to the one mentioned above.
Basel Precision Instruments SP983c-IF current preamplifiers and SP1004 voltage preamplifiers were used to achieve lower measurement noise.
A Yokogawa GS210 was used to apply DC currents to devices through a \SI{10}{\mega\ohm} bias resistor.
Other electronic instruments used in dilution refrigerator measurements are the same as those used in device pre-characterization.
The temperature was measured using a calibrated RX-102A thermometer mounted on the sample probe.
The temperature was controlled using a probe heater mounted close to both the thermometer and the thermal anchors of the wires, to guarantee good thermal connection among devices, the thermometer, and the heater.

The remaining dilution refrigerator measurements were performed in a wet Oxford Kelvinox TLM dilution refrigerator.
Thermocoax cables were used from room temperature to $\SI{4}{K}$, followed by RC filters before connecting to the sample.
DC voltages were applied to the top and bottom gates using Yokogawa 7651 voltage sources.
The AC currents and voltages were preamplified by DL-1211 current preamplifiers and SR560 voltage preamplifiers respectively, then measured by SR830 lock-in amplifiers.

High magnetic field measurements (up to $\SI{18}{T}$) were taken using SCM-1 dilution refrigerator at the National High Magnetic Field Laboratory, with an $\SI{18}{T}$ superconducting magnet and a base temperature of about $\SI{20}{mK}$.
A home-made $\SI{20}{cm}$ twisted-pair copper tape filter was installed and immersed in the helium-3/helium-4 mixture to facilitate thermalization and filter high-frequency Johnson noise.
The aforementioned electronics were used in the measurements.

By controlling voltages applied to the top and bottom gates ($V_\text{tg}$ and $V_\text{bg}$), the electron density, $n$, and the perpendicular electric displacement field, $D$, can be tuned independently, following relations: $n=(\epsilon_\mathrm{BN}\epsilon_0/e$)($V_\mathrm{bg}$/$d_\mathrm{bg} + V_\mathrm{tg}$/$d_\mathrm{tg}$) and $D=\pm(\epsilon_\mathrm{BN}\epsilon_0/2$)($V_\mathrm{bg}$/$d_\mathrm{bg} - V_\mathrm{tg}$/$d_\mathrm{tg}$).
Here, $\epsilon_\mathrm{BN}=3$ is the relative dielectric constant of hBN, $\epsilon_0$ is the vacuum permittivity, $e$ is the elementary charge, and $d_\mathrm{bg}$ ($d_\mathrm{tg}$) is the thickness of the bottom (top) hBN.
In this work, we defined $D$ such that $D>0$ corresponds to a displacement field oriented from the small twist angle \moire towards the large twist angle \moire.

\begin{figure*}
    \centering
    \includegraphics[width=3.5in]{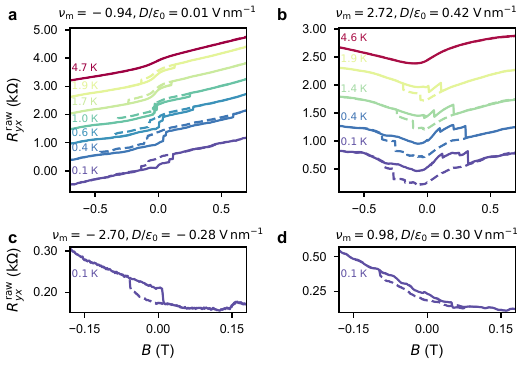}
    \caption{\figtitle{Field sweep measurements raw data.}
    \panel{a-d}, same as \figlbls~\ref{fig:2}\panel{d-g}, only showing raw data $R_{yx}$, without antisymmetrization in magnetic field.
    \label{fig:AHE_raw}
    }
\end{figure*}

To map the anomalous Hall response shown in \figlbls~\ref{fig:2}\panel{b, c} and minimize the contribution from both the regular Hall effect and $R_{xx}$ mixing due to anisotropy, we measured $R_{yx}$ at $B=\pm\SI{60}{mT}$ and $\pm\SI{30}{mT}$.
The anomalous Hall response $R_{yx}^\mathrm{AHE}$ is then calculated according to $R_{yx}^\mathrm{AHE}=(R_{yx}^{\SI{30}{mT}}-R_{yx}^{-\SI{30}{mT}})-(R_{yx}^{\SI{60}{mT}}-R_{yx}^{-\SI{60}{mT}})/2$.
This procedure removes the regular Hall component when its contribution to $R_{yx}$ has a linear dependence on $B$, which is in general true for the small $B$ considered here.
For the field sweep measurements shown in \figlbls~\ref{fig:2}\panel{d-g}, $R_{yx}$ is antisymmetrized between curves of opposite sweep direction, so that $R_{yx}=(R_{yx}^\mathrm{raw \uparrow}(B)-R_{yx}^\mathrm{raw \downarrow}(-B))/2$ (see $R_{yx}^\mathrm{raw}(B)$ in \extlbl~\ref{fig:AHE_raw}).
Here, the arrows indicate the sweep direction of $B$.

\subsection{Twist angle determination}\label{ssec:twist}

We extract the interlayer twist angles of TTG devices through the following procedure: 
(1) By fitting the Landau levels to the Str\v{e}da formula with the corresponding Chern numbers as extracted according to the quantized values of $R_{yx}$, we can calibrate the top and bottom gate capacitances.
(2) Using these capacitance values, the charge density corresponding to $\nu_\mathrm{m}=\pm 4$ can be identified using $R_{xx}$ peaks at integer fillings and Landau levels emanating from the charge neutral point and the band extrema.
(3) The smaller twist angle can be calculated: $n_{\nu=\pm4} = \pm 8\sin^{2}\theta_{12}/\sqrt{3}a_0^2 \approx \pm 8\theta_{12}^2/\sqrt{3}a_0^2$.
The errors in $\theta_{12}$ depend on the uncertainty in the fits to $R_{xx}$ peaks and Landau levels, which are less than $0.05\degree$ in general.

\begin{figure*}
    \centering
    \includegraphics[width=9cm]{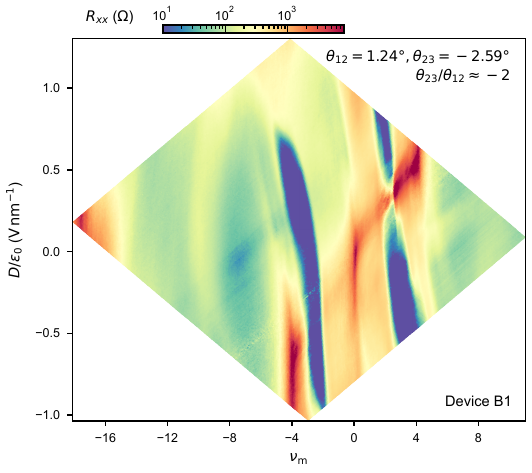}
    \caption{\figtitle{$R_{xx}$ versus $\nu_\mathrm{m}$ and $D$ measured for Device B1 with extended gate ranges.}
    Same measurement is presented in \figlbl~\ref{fig:3}\panel{a}.
    From the $R_{xx}$ peak at $\nu_\mathrm{m}=-17.5$, we extract $|\theta_{23}|=2.59\degree$.
    \label{fig:AH05_t23}
    }
\end{figure*}

The determination of the larger angle, $\theta_{23}$, is more challenging given that the breakdown voltage of the hBN dielectric usually does not allow us to access $\nu_{\pm4}$ fillings for layers 2 and 3.
Here, we comment on this issue for each device.
For Device A, for which we targeted $\theta_{23}/\theta_{12}=2$, the fact that the AHE is observed indicates that the system relaxes to \moire periodic domains with the local angle ratio being exactly $2$ (see \figlbl~\ref{fig:2}\panel{h}).
When the global angle ratio deviates from $2$, these domains are still expected to form with smaller sizes as long as the deviation is small \cite{Xia2025Topological}.
Therefore, the physical properties of the system remain qualitatively the same regardless of the exact value of $\theta_{23}$.
For Device B1, for which we targeted $\theta_{23}/\theta_{12}=-2$,  we are able to dope to 4 holes per \moire unit cell of layer 2 and 3 (see results with extended gate ranges in \extlbl~\ref{fig:AH05_t23}).
From its density, we extract $\left|\theta_{23}\right|=2.59\degree$ for Device B1.
The extracted ratio between the two twist angles is $-2.09$, close to the target value.
For Device B2, Device C, and Device D, we targeted ratios of $\theta_{23}/\theta_{12}=-2, -3$ and $3$, respectively.
The fact that we did not observe any $R_{xx}$ peaks corresponding to the \supermoire density~\cite{Xie2025Strong} sets an upper bound on the \supermoire density, and thus a lower bound on the \supermoire wavelength.
From the width of the $R_{xx}$ peak at charge neutrality, we estimate $l_\mathrm{sm}/l_\mathrm{m}\gtrsim 7$ for the devices studied here.
From this estimated minimum $l_\mathrm{sm}$, we can calculate the range of $\theta_{23}$ for each device, as shown in \extlbls~\ref{fig:SM_xi}\panel{c-e}.
The resultant twist angle ratio ranges are $[-2.14, -1.86]$, $[-3.13, -2.87]$, and $[2.92, 3.07]$, for Device B2, Device C, and Device D, respectively.
Although the exact values of $\theta_{23}$ are unknown for these devices, we argue that all discussions and conclusions presented in the paper are valid within the estimated $\theta_{23}$ ranges, since the main effect will be on the size of $l_\mathrm{sm}$ (see Methods~\ref{ssec:SM}).
For Device E and F, we targeted an interlayer twist between layers 2 and 3 of $15\degree$.
Within the possible error range, layer 3 will be electronically decoupled, representing the asymptotic limit of both branches of the magic continuum.
Therefore, we argue that the exact value of $\theta_{23}$ is not important in these cases.

\subsection{Correlated states of TTG \moire quasicrystals in finite magnetic fields}\label{ssec:Landau_fan}

\begin{figure*}
    \centering
    \includegraphics[width=7in]{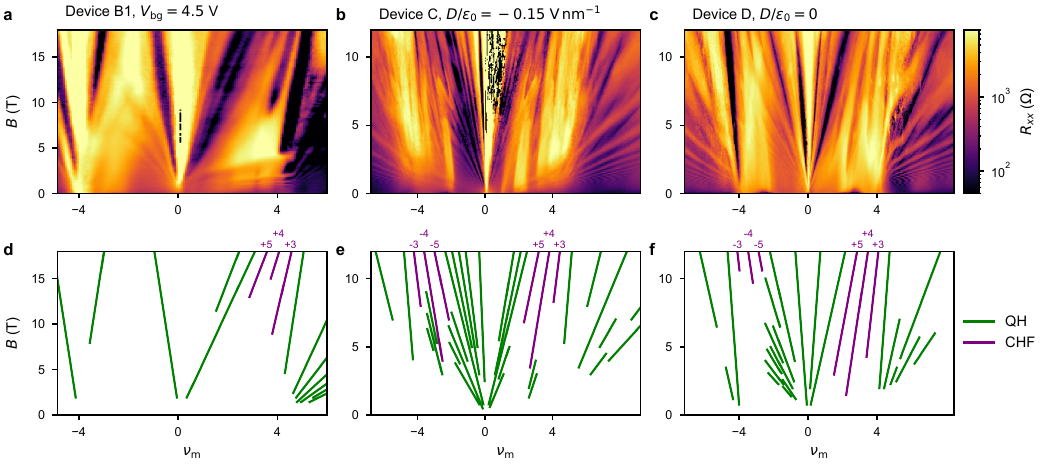}
    \caption{\figtitle{Landau fan diagram of quasicrystalline TTG.}
    \panel{a-c}, $R_{xx}$ versus $\nu_\mathrm{m}$ and $B$ measured in quasicrystalline TTG devices shown in \figlbl~\ref{fig:3}, while keeping $V_\mathrm{bg}=\SI{4.5}{V}$, $D/\epsilon_0=-\SI{0.15}{V\,nm^{-1}}$, and $D/\epsilon_0=0$, respectively.
    All data is taken in dilution refrigerators with mixing chamber temperature lower than $\SI{30}{mK}$.
    \panel{d-f}, Labels of main features in Landau fan diagrams shown in \panel{a-c}, which are categorized into quantum hall states (QH) and correlated Hofstadter ferromagnetic states (CHF).
    \label{fig:Landau_fan}
    }
\end{figure*}

Extended~Data~Figure~\ref{fig:Landau_fan} shows Landau fan diagrams of our quasicrystalline TTG devices.
Apart from quantum Hall states emanating from band edges and the charge neutrality point, the most prominent features that develop under an applied magnetic field exhibit Chern numbers $C=\pm 5, \pm 4, \pm 3$, as deduced from their Str\v{e}da $n$-$B$ slopes.
These features extrapolate to a zero-field band fillings of $\nu_\mathrm{m}=\pm 1, \pm 2, \pm 3$, respectively.
A similar sequence of correlated Chern insulators has been reported in magic-angle twisted bilayer graphene~\cite{Tomarken2019Electronic,Nuckolls2020Strongly,Choi2021Correlation-driven,Park2021Flavour,Wu2021Chern,Saito2021Hofstadter,Das2021Symmetry-broken,Stepanov2021Competing,Yu2022Correlated,He2025Strongly,Finney2025Extended}, the microscopic nature of which is likely the correlated Hofstadter ferromagnets~\cite{Wang2024Theory}.

\subsection{Further discussion of \supermoire-modulated superconductivity}\label{ssec:SM_SC}

\begin{figure*}
    \centering
    \includegraphics[width=14cm]{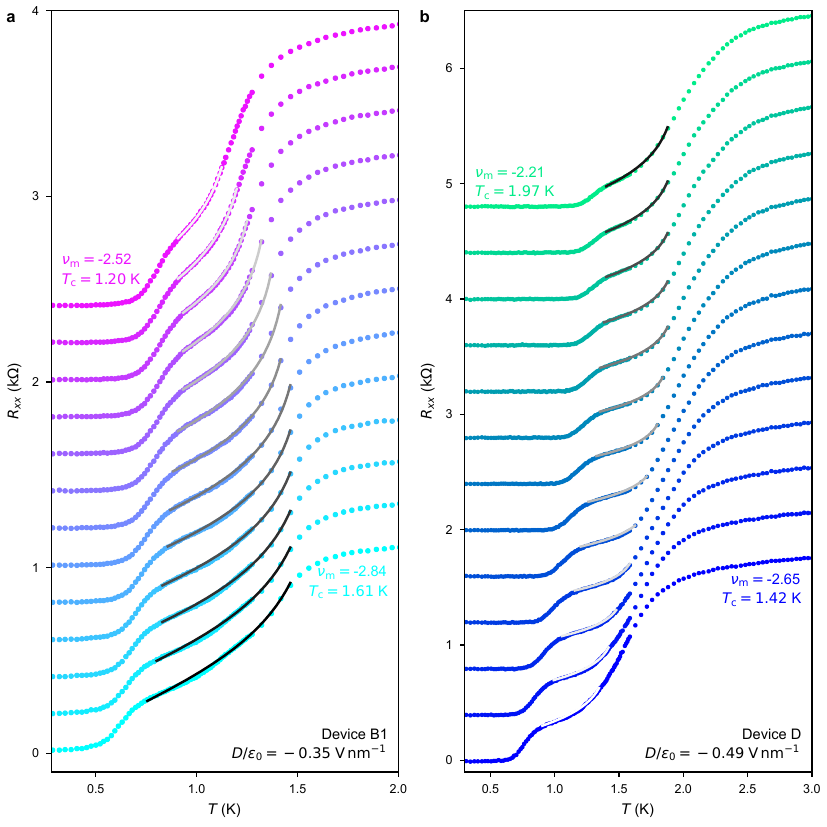}
    \caption{\figtitle{Superconducting proximity model fitting of $R_{xx}(T)$ traces between two transitions.}
    \panel{a}, $R_{xx}(T)$ traces reproduced from \figlbl~\ref{fig:4}\panel{a}.
    Data taken at different $\nu_\mathrm{m}$ is shifted by $\SI{0.2}{k\Omega}$.
    \panel{b}, $R_{xx}(T)$ traces reproduced from \figlbl~\ref{fig:4}\panel{b}.
    Data taken at different $\nu_\mathrm{m}$ is shifted by $\SI{0.4}{k\Omega}$.
    Solid curves show the fitting results from the superconducting proximity model.
    The $T_\mathrm{c}$ values obtained from the fits are indicated for the first and last curves.
    \label{fig:RTfit}
    }
\end{figure*}

The superconducting proximity model~\cite{Abraham1982Resistive} gives the following temperature dependence of $R_{xx}$ between the two superconducting transitions, which we use to fit the experimental data:
\begin{equation}
R_{xx}(T) = R_{xx}^0\left(1-A\xi_n\ln\left(\frac{B\xi_n(1-T/T_\mathrm{c})^2}{T}\right)\right).
\end{equation}
Here, $R_{xx}^0$ is the sample resistance after the islands become superconducting but before the proximity effect begins to reduce the resistance, $T_\mathrm{c}$ is the superconducting transition temperature of the islands, and $\xi_n$ is the normal metal coherence length.
Here, we neglect the temperature dependence of $\xi_n$ and treat it as a fitting parameter, same as $A$ and $B$.
The fitting results are shown in \extlbl~\ref{fig:RTfit}.
This model does not consider the BKT phase transition and therefore is not a meaningful fit of the experimental data below the lower transition.
The extracted values of $T_\mathrm{c}$ match the higher transition temperatures of $R_{xx}(T)$ traces.

\begin{figure*}
    \centering
    \includegraphics[width=9cm]{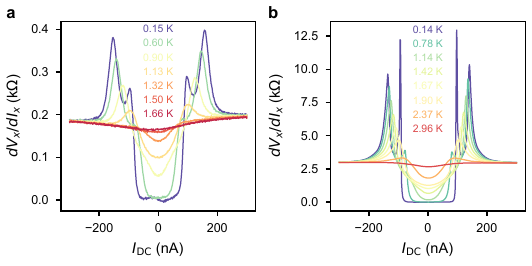}
    \caption{\figtitle{Temperature dependent differential resistance across superconducting transitions.}
    \panel{a,b}, $dV_{x}/dI_{x}$ versus $I_\mathrm{DC}$ measured at different $T$.
    Data is reproduced from \figlbls~\ref{fig:4}\panel{a,b}.
    \label{fig:IVcut}
    }
\end{figure*}

\begin{figure*}
    \centering
    \includegraphics[width=9cm]{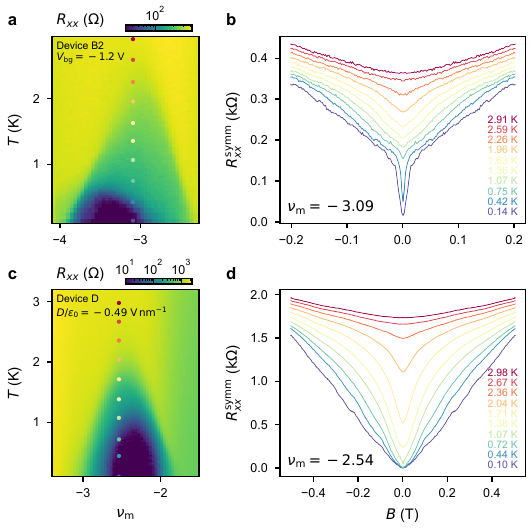}
    \caption{\figtitle{Temperature dependent magnetoresistance across superconducting transitions.}
    \panel{a,c}, $R_{xx}$ versus $\nu_\mathrm{m}$ and $T$ measured for Device B2 at $V_\mathrm{bg}=-\SI{1.2}{V}$ (\panel{a}) and Device D at $D/\epsilon_0=-\SI{0.49}{V\,nm^{-1}}$ (\panel{c}).
    Measurements under the same conditions are also presented in \extlbl~\ref{fig:2dome}\panel{e} left and \figlbl~\ref{fig:3}\panel{f} left, respectively.
    \panel{c} and \figlbl~\ref{fig:3}\panel{f} left are not identical, because they were taken in different measurement sessions and Device D changed slightly during the thermal cycle.
    However, key properties of the sample, including correlated states and robust superconductivity with the two-step transition behaviour, still remain.
    \panel{b,d}, Magnetoresistance measured at different $T$.
    The measurement conditions for each curves shown in \panel{b} (\panel{d}) are indicated by dots with the same colour code in \panel{a} (\panel{c}).
    Between the two transitions, both samples demonstrate a linear magnetoresistance at small $B$, consistent with the behaviour of Josephson junction arrays with free-moving vortices.
    Here, to remove the contribution from $R_{yx}$ mixing, we plot symmetrized magnetoresistance $R_{xx}^\mathrm{symm}$, which is calculated from the the measured $R_{xx}$ using the relation $R_{xx}^\mathrm{symm}\left(B\right)=\left(R_{xx}\left(B\right)+R_{xx}\left(-B\right)\right)/2$.
    \label{fig:MR}
    }
\end{figure*}

To better understand the nature of the two transitions, we measured differential resistance $dV_{x}/dI_{x}$ as a function of $I_\mathrm{DC}$ and $T$, shown in \figlbls~\ref{fig:4}\panel{c,d} and \extlbl~\ref{fig:IVcut}.
At base temperature, two critical currents can be identified as peaks of $dV_{x}/dI_{x}$.
Between them, non-zero $dV_{x}/dI_{x}$ is observed that is consistent with the flux-flow regime of the Josephson junction array~\cite{Orlando1991Phenomenological}.
This phase evolves towards lower $I_\mathrm{DC}$ with increasing temperature and exists at $I_\mathrm{DC}=0$ between the two transition temperatures, suggesting that part of the sample remains superconducting in this temperature range.
This conclusion is corroborated by the measurements of $R_{xx}(B)$ at different $T$, shown in \extlbl~\ref{fig:MR}.
At temperatures above the higher transition, the increase of $R_{xx}$ is proportional to $B^2$, as expected from the Drude model~\cite{Pippard1989Magnetoresistance}.
At temperatures between the two transitions, we observe a linear magnetoresistance at small $B$, consistent with the behaviour of Josephson junction arrays at small vortex densities~\cite{Rzchowski1990Vortex}.

For \supermoire-modulated superconductivity to emerge, two conditions must be satisfied: (1) $l_\mathrm{m}\ll l_\mathrm{sm}$ so that the \supermoire potential can be viewed as a slow modulation of a local electronic structure, and (2) $\xi_\mathrm{GL}\lesssim l_\mathrm{SC}<l_\mathrm{sm}$ so that superconductivity can be established within certain regions in the \supermoire unit cell.
Here, $\xi_\mathrm{GL}$ is the Ginzburg-Landau coherence length, and $l_\mathrm{SC}$ is the size of the superconducting islands between two transitions.
The first condition is supported by the absence of $R_{xx}$ peaks at the \supermoire density~\cite{Xie2025Strong}, from which we estimate $l_\mathrm{sm}/l_\mathrm{m}\gtrsim 7$ for the devices studied here as discussed in Methods~\ref{ssec:twist}.
To verify the second condition, we compare the $l_\mathrm{sm}$ and $\xi_\mathrm{GL}$ that we extract from the dependence of $R_{xx}$ on $B$ and $T$ (see \extlbl~\ref{fig:SM_xi}, Methods~\ref{ssec:SM}, and Methods~\ref{ssec:xi}).
$\xi_\mathrm{GL}$ is about $\SI{15}{nm}-\SI{50}{nm}$ across different devices, less than the expected \supermoire wavelength, $l_\mathrm{sm}\gtrsim \SI{100}{nm}$, for devices with $\theta_{23}/\theta_{12}\approx -2$ or $3$.

It is worth noting that we did not observe oscillations of $R_{xx}$ when the number of flux quanta per \supermoire unit cell is an integer or a simple fraction, as reported in other Josephson junction array systems~\cite{Tinkham1983Periodic,Bøttcher2018Superconducting}.
Possible reasons include considerable \supermoire disorder arising from twist angle variations and strain, as well as finite size effects.
Future investigations will benefit from advances in fabrication techniques that minimize \supermoire disorder, as well as from scanning probes with \moire- or atomic-scale spatial resolution, which are essential for further studying the effects of \supermoire on superconductivity.
In Device C, where $\theta_{23}/\theta_{12} \approx -3$, the signatures of \supermoire-modulated superconductivity are weak or absent within the superconducting phase space.
This could originate from the weaker $\mathbf{d}$-dependence of the local electronic structure, as illustrated in \figlbl~\ref{fig:1}\panel{h}, suggesting that the third graphene layer is weakly coupled in this system.

\subsection{\Supermoire wavelength calculations}\label{ssec:SM}

When the ratio between the two twist angles of a TTG heterostructure is close to a simple fraction, i.e., $\theta_{12}/\theta_{23}\approx p/q$ with $p$ and $q$ being small coprime integers, the resulting structure can be understood approximately in terms of a commensurate \moire unit cell (consisting of $p\times p$ and $q\times q$ bilayer \moire unit cells defined by $\theta_{12}$ and $\theta_{23}$, respectively), but with the offset between the two \moire patterns varying slowly at a \supermoire length scale given by
\begin{align}
l_\mathrm{sm} = & \frac{a_0}{\sqrt{2q^2\left(1-\cos\theta_{12}\right)+2p^2\left(1-\cos\theta_{23}\right)-4\left|pq\right|\cos\left(\frac{\theta_{12}+\theta_{23}}{2}\right)\sqrt{\left(1-\cos\theta_{12}\right)\left(1-\cos\theta_{23}\right)}}}.\label{eqn:lsm}
\end{align}
When both $\left|\theta_{12}\right|$ and $\left|\theta_{23}\right|$ are small and $\theta_{12}/\theta_{23}=p/q$, this equation can be approximated as $l_\mathrm{sm}\approx2a_0/\left|\left(p+q\right)\theta_{12}\theta_{23}\right|$.
In practice, we calculate $l_\mathrm{sm}$ for arbitrary twist angle combinations by evaluating Eq.~\ref{eqn:lsm} for $1\leq |p|, |q|\leq 10$ and choosing the largest resulting value (\extlbls~\ref{fig:SM_xi}\panel{a, b}).

\subsection{Extraction of \texorpdfstring{$\xi_\mathrm{GL}$}{TEXT}}\label{ssec:xi}

\begin{figure*}
    \centering
    \includegraphics[width=14cm]{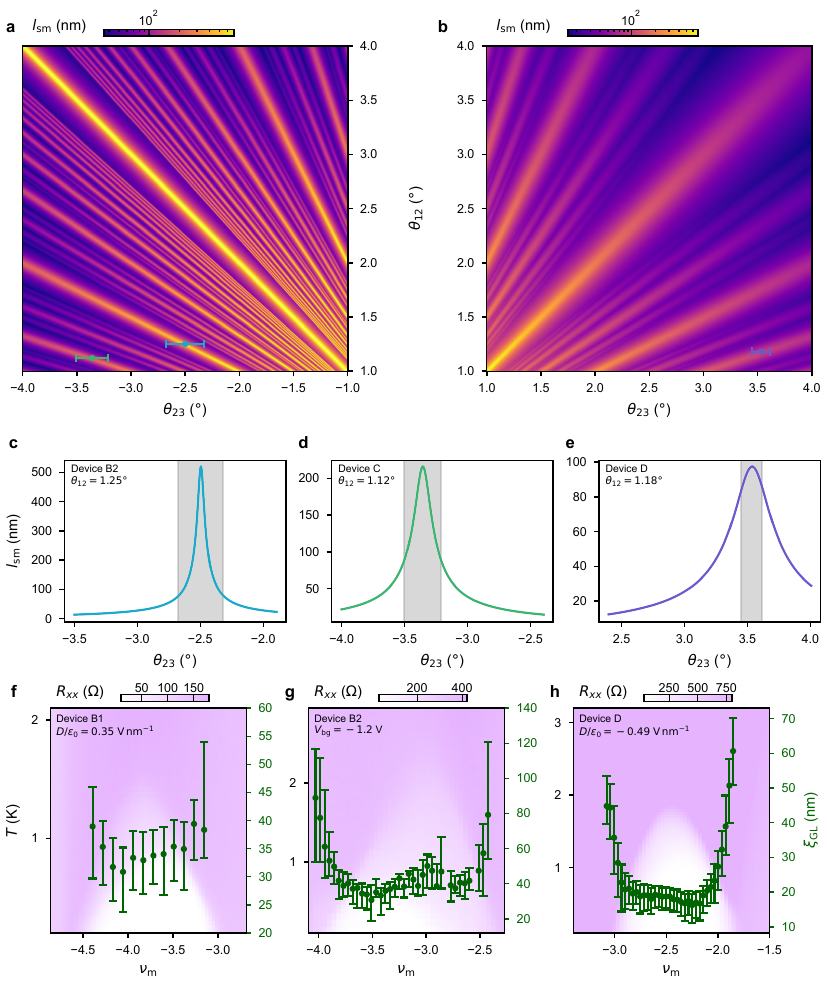}
    \caption{\figtitle{Comparison between \supermoire wavelength and Ginzburg-Landau coherence length.}
    \panel{a,b}, Calculated \supermoire wavelength as a function of $\theta_{12}$ and $\theta_{23}$ for alternating-twisted TTG (\panel{a}) and helically-twisted TTG (\panel{b}).
    See Methods~\ref{ssec:SM} for calculation details.
    \panel{c-e}, $l_\mathrm{sm}$ calculated using Eq.~\ref{eqn:lsm} as a function of $\theta_{23}$ for Device B2 (\panel{c}), Device C (\panel{d}), and Device D (\panel{e}).
    Here, we use the experimental values of $\theta_{12}$ and targeted $(p, q)$ in Eq.~\ref{eqn:lsm}.
    Grey-shaded regions show the $\theta_{23}$ ranges for each device, which are estimated from the absence of $R_{xx}$ peaks corresponding to the \supermoire density (see Methods~\ref{ssec:twist} for details).
    These devices are marked in \panel{a,b} using the same colour code, where the error bars are the estimated $\theta_{23}$ range.
    \panel{f-h}, Extracted Ginzburg-Landau coherence length $\xi_\mathrm{GL}$ versus $\nu_\mathrm{m}$ for Device B1 (\panel{f}), Device B2 (\panel{g}), and Device D (\panel{h}).
    Colour maps show $R_{xx}$ versus $\nu_\mathrm{m}$ and $T$, reproduced from \extlbl~\ref{fig:2dome}\panel{d} left (\panel{f}), \extlbl~\ref{fig:2dome}\panel{e} left (\panel{g}), and \extlbl~\ref{fig:MR}\panel{c} (\panel{h}).
    See Methods~\ref{ssec:xi} for details of $\xi_\mathrm{GL}$ extraction.
    \label{fig:SM_xi}
    }
\end{figure*}

To extract $\xi_\mathrm{GL}$, we analyse the magnetic field dependence of the superconducting transition temperature.
For each magnetic field, the $R_{xx}(T)$ trace above the higher transition is first fitted with a linear function of $T$ to approximate the normal-state resistance.
The superconducting transition temperature $T_c$ is then defined as the temperature at which the measured resistance curve intersects 85\%, 90\%, or 95\% of the fitted curve.
The extracted $T_c$ values are fitted as a linear function of $B$, following the Ginzburg-Landau relation:
\begin{equation}
\frac{T_c}{T_c^0}=1-\frac{2\pi\xi_\mathrm{GL}^2}{\Phi_0}B,
\end{equation}
where $\Phi_0=h/2e$ is the superconducting flux quantum, and $T_c^0$ is the zero-field superconducting transition temperature.
The slope of the linear fit determines $\xi_\mathrm{GL}$.
While $\xi_\mathrm{GL}$ values extracted using different resistance thresholds (85\%, 90\%, 95\%) differ slightly, they exhibit consistent trends as a function of $\nu_\mathrm{m}$.
In \extlbl~\ref{fig:SM_xi}, we plot the 90\% criterion value of $\xi_\mathrm{GL}$ as the data point, with the range defined by the 85\% and 95\% criteria shown as error bars.
Since the extraction is based on the field dependence of the higher transition, the obtained coherence length characterizes the superconducting regions that exist between the two transitions.

\subsection{Electronic structure calculations}\label{ssec:calc}

In this section, we describe the electronic band structure calculation of $\theta_{12}/\theta_{23}\approx p/q$ TTG shown in \figlbl~\ref{fig:1}, which are calculated using a continuum model for a commensurate $p/q$ \moire supercell.

We start with the atomic lattice vectors for each graphene layer, given by the columns of the matrix $\bm{A}_{\ell}=\bm{R}(\theta_\ell)\bm{A}_0$, where
$\ell=1,2,3$ labels the layer index, $(\theta_1,\theta_2,\theta_3)=(-p\theta_0,0,q\theta_0)$ are the twist angles of the three layers, and
\begin{equation}
\bm{R}(\theta)=\begin{pmatrix}
\cos\theta & -\sin\theta\\
\sin\theta & \cos\theta
\end{pmatrix} ;\;\;\;\;
\bm{A}_0=
a_0\begin{pmatrix}
1 & \frac{1}{2} \\
0 & \frac{\sqrt{3}}{2}
\end{pmatrix}
\end{equation}
with $a_0=\SI{0.246}{nm}$.
The \moire superlattice vectors are given by the columns of the matrix $\bm{A}_{12}\equiv (\bm{A}_1^{-1}-\bm{A}_2^{-1})^{-1}$ and $\bm{A}_{23}\equiv (\bm{A}_2^{-1}-\bm{A}_3^{-1})^{-1}$.
In general, these two \moire superlattices are incommensurate but, for small twist angles, are almost commensurate $p\bm{A}_{12}\approx q\bm{A}_{23}$.

To proceed, we construct a local periodic structure.
We consider a slight distortion of the lattices, which we denote with a prime $\bm{A}_{\ell}^\prime$, for which the \moire scale structure is exactly periodic.
Specifically, we take $\bm{A}_{1}^\prime=\bm{A}_1$ and $\bm{A}_3^\prime=\bm{A}_3$ to be unchanged, but slightly deform the middle layer as
\begin{equation}
\bm{A}_{2}\rightarrow\bm{A}_2^\prime=\left(\frac{q}{p+q}\bm{A}_1^{-1}+\frac{p}{p+q}\bm{A}_3^{-1}\right)^{-1}
\end{equation}
which results in the new \moire superlattice vectors satisfying
\begin{equation}
p\bm{A}_{12}^\prime=q\bm{A}_{23}^\prime=(p+q)\left(\bm{A}_1^{-1}-\bm{A}_3^{-1}\right)^{-1} \equiv\bm{A}_{m}
\end{equation}
thus defining a commensurate \moire unit cell $\bm{A}_m$.
To see that this is a small modification of the actual structure at small twist angles, notice that the deviation
\begin{equation}
\bm{A}_2^\prime - \bm{A}_2 = 
\frac{2pq}{p+q}i\theta_0\sigma_y\bm{A}_0 + O(\theta_0^2)
\end{equation}
is only non-zero at order $\theta_0$ (in this analysis, we have assumed $p/q\neq -1$).

We can now calculate the local electronic properties.
We consider three graphene sheets in the commensurate structure.
Importantly, we allow each layer to be displaced in-plane according to a vector $\bm{s}_{\ell}$, such that the sublattice $\sigma\in \{A,B\}$ carbon atoms in layer $\ell$ are positioned at $\bm{A}_{\ell}^\prime[(n,m)^{\mathsf{T}}+\bm{t}_{\sigma}+\bm{s}_{\ell}]$, where $\bm{t}_{A}=0$ and $\bm{t}_{B}=(2/3,1/3)^{\mathsf{T}}$.
This shift is important to capture the local stacking configurations of trilayer graphene, which leads to different electronic properties.

We model the electronic properties using the standard continuum model approach, which is valid at small twist angles.
We work in the Hilbert space spanned by $\{|\bm{k},\ell,\sigma\rangle\}$, where $\bm{k}$ is momentum, $\ell=1,2,3$ is layer, and $\sigma=A,B$ is sublattice.
Let $\bm{B}_{\ell}^\prime = 2\pi (\bm{A}_{\ell}^\prime)^{-\mathsf{T}}$.
We take $\bm{k}$ to be near the $\bm{K}\equiv\bm{B}_{0}(2/3,1/3)^{\mathsf{T}}$ point.
The intralayer terms are given by the standard Dirac equation near the $\bm{K}_{\ell}=\bm{B}_{\ell}^\prime(2/3,1/3)^{\mathsf{T}}$ point,
\begin{equation}
\langle \bm{k},\ell,\sigma^\prime|H|\bm{k},\ell,\sigma\rangle = v_F\begin{pmatrix}
0 & e^{i\theta_\ell}[k_x - i k_y - (K_{\ell,x}-iK_{\ell,y})] \\
c.c. & 0
\end{pmatrix}_{\sigma^\prime \sigma}
\end{equation}
For the interlayer tunnelling terms, we keep the (most dominant) first harmonic terms.  
For adjacent layers $|\ell^\prime-\ell|=1$,
the non-zero interlayer tunnelling matrix elements are given by
\begin{equation}
\langle \bm{k}+\bm{q}_n^{\ell^\prime \ell},\ell^\prime,\sigma^\prime|H|\bm{k},\ell,\sigma\rangle = T^{\ell^\prime \ell}_{n\sigma^\prime\sigma}(\bm{k})
\end{equation}
for $n=0,1,2$, indexes the three tunnelling matrix elements that shift momentum by $\bm{q}_n^{\ell^\prime\ell}=(\bm{B}_{\ell^\prime}^\prime-\bm{B}^\prime_{\ell})\bm{z}_n$, with $\bm{z}_0=(0,0)^\mathsf{T}$, $\bm{z}_1=(-1,0)^\mathsf{T}$, $\bm{z}_2=(-1,-1)^\mathsf{T}$.
The matrix elements are given by
\begin{equation}
T^{\ell^\prime \ell}_{n\sigma^\prime\sigma}(\bm{k}) = 
we^{2\pi i \bm{z}_n\cdot(\bm{s}_{\ell}-\bm{s}_{\ell^\prime})}(1+\xi(|\bm{k}+\bm{B}_{\ell}\bm{z}_n|-|\bm{K}|))
\begin{pmatrix}
1 & \kappa e^{-2\pi i n / 3} \\
\kappa e^{2\pi i n /3} & 1
\end{pmatrix}_{\sigma^\prime \sigma}
\end{equation}
There are a few differences in comparison to the standard Bistritzer-MacDonald for twisted bilayer graphene~\cite{Bistritzer2011Moire}.  
First, we explicitly take the in-plane displacements $\bm{s}_{\ell}$ into account via the phase factor $e^{2\pi i \bm{z}_n \cdot (\bm{s}_{\ell}-\bm{s}_{\ell^\prime})}$.
While for a bilayer, such a shift amounts to a redefinition of the origin, in the commensurate trilayer this shift is physical and impacts the electronic band structure.
Second, we keep the first-order correction in the momentum dependence of the tunnelling term via the $\xi$ term~\cite{Xia2025Topological,Kwan2024Fractional,Hoke2024Imaging}.
We use the parameters $v_F=0.88\times 10^6$ $\mathrm{m\,s^{-1}}$, $w=\SI{0.11}{eV}$, $\xi=\SI{-2.1}{\text{\AA}}$, and $\kappa=0.68$.

In terms of the atomic displacements $\bm{s}_\ell$, the vector connecting the AA$_{12}$ and AA$_{23}$ sites in the commensurate structure is~\cite{Yang2024Multi}
\begin{equation}
\bm{d}=\bm{A}_m\left(\frac{1}{p}(\bm{s}_1-\bm{s}_2)-\frac{1}{q}(\bm{s}_2-\bm{s}_3)\right)
\end{equation}
The electronic structure only depends on $\bm{s}_{\ell}$ through $\bm{d}$ modulo $\frac{1}{|pq|}\bm{A}_m$.  
We take $\bm{s}_2=\bm{s}_3=(0,0)^{\mathsf{T}}$ and set
$\bm{s}_1=(0,0)^{\mathsf{T}}$ for the $\bm{d}=0$ stacking, $\bm{s}_1=\frac{1}{|q|}(1/3,1/3)^{\mathsf{T}}$ for the $\bm{d}=\bm{\delta}$ stacking, and $\bm{s}_1=\frac{1}{|q|}(1/2,0)^{\mathsf{T}}$ for the $\bm{d}=\bm{\beta}$ stacking configurations.

\section*{Acknowledgements}
We thank C. Yang for sharing unpublished calculation results.
We thank B. Feldman for helpful feedback on the paper.
We thank G. de la Fuente Simarro for assistance with device fabrication and transport measurements.
We thank A. Bangura, G. Jones, R. Nowell, A. Woods, and S. Hannahs for supporting measurements performed at the National High Magnetic Field Laboratory.
Work in the P.J.-H. group was partially supported by the Army Research Office MURI W911NF2120147, the Air Force Office of Scientific Research (AFOSR) grant FA9550-21-1-0319, the National Science Foundation (DMR-1809802), a Max Planck-Humboldt Research Award to P.J.-H., the CIFAR Quantum Materials Program, the Ramón Areces Foundation, and the Gordon and Betty Moore Foundation’s EPiQS Initiative through grant GBMF9463 to P.J.-H.
A.U. acknowledges support from the MIT Pappalardo Fellowship and from the VATAT Outstanding Postdoctoral Fellowship in Quantum Science and Technology.
A.S. was supported by the US Department of Energy, Office of Science, Basic Energy Sciences, Materials Sciences and Engineering Division, under Contract DE-AC02-76SF00515.
F.G. is grateful for the financial support from the Swiss National Science Foundation (Postdoc.Mobility Grant No. 222230).
K.W. and T.T. acknowledge support from the JSPS KAKENHI (Grant Numbers 21H05233 and 23H02052) , the CREST (JPMJCR24A5), JST and World Premier International Research Center Initiative (WPI), MEXT, Japan.
L.F. was supported primarily by Simons Investigator Award from Simons Foundation.
T.D. was supported by a startup fund at Stanford University.
J.H.S. acknowledges financial support from the German Science Foundation through the SPP2244 program.
A portion of this work was performed at the National High Magnetic Field Laboratory, which is supported by National Science Foundation Cooperative Agreement No. DMR-2128556 and the State of Florida.
This work was performed in part at the Harvard University Center for Nanoscale Systems (CNS); a member of the National Nanotechnology Coordinated Infrastructure Network (NNCI), which is supported by the National Science Foundation under NSF award no. ECCS-2025158.
This work was carried out in part through the use of MIT.nano's facilities.

\section*{Author contributions}

A.U., L.-Q.X., and P.J.-H. conceived the project. L.-Q.X. fabricated the devices with the help of A.U. L.-Q.X. carried out the helium-3 refrigerator measurements with the help of A.U., J.Y., and A.S. L.-Q.X. and A.U. carried out a portion of the dilution refrigerator measurements under the supervision of P.J.-H. J.Y. carried out the remaining dilution refrigerator measurements under the supervision of J.H.S. L.-Q.X. and A.S. carried out measurements at the National High Magnetic Field Laboratory. T.D., N.S.T., and J.M.-M. performed band structure and lattice relaxation calculations. K.W. and T.T. supplied the boron nitride crystals. L.-Q.X., A.U., J.Y., A.S., F.G., N.S.T., J.M.-M., T.D., L.F., and P.J.-H. analysed the data and discussed the interpretation. L.-Q.X. wrote the manuscript with the help of A.U. and A.S., and input from all authors. P.J.-H. supervised the project.

\section*{Competing interests}

The authors declare no competing interests.

\section*{Data availability}

The data that support the findings of this study are available from the corresponding authors upon reasonable request.

\bibliography{ref}

\clearpage
\ExtendedData

\begin{table*}
    \begin{booktabs}{cccc}
    \toprule
    \textbf{Device} & \textbf{Approximate} $\theta_{23}/\theta_{12}$ & $\nu_\mathrm{m}$ \textbf{of correlated states} & \textbf{Corresponding figure} \\
    \midrule[solid]
    Device A & 2 & $-1, -2, -3$ & \figlbl~\ref{fig:2}\panel{a} \\
    \midrule[dotted]
    Device B1 & -2 & $8/3$ & \figlbl~\ref{fig:3}\panel{a} \\
    \midrule[dotted]
    Device B2 & -2 & $-2, 1, 2, 3$ & \extlbl~\ref{fig:2dome}\panel{b} \\
    \midrule[dotted]
    Device C & -3 & $-2, 1, 2, 3$ & \figlbl~\ref{fig:3}\panel{b} \\
    \midrule[dotted]
    Device D & 3 & $-2, 1, 2, 3$ & \extlbls~\ref{fig:ML11_highT}\panel{a,b} \\
    \midrule[dotted]
    Device E & $>5$ & $-2, 1, 2, 3$ & \extlbl~\ref{fig:MATBG+MLG}\panel{a} \\
    \midrule[dotted]
    Device F & $<-5$ & $-3, -2, 2, 3$ & \extlbl~\ref{fig:MATBG+MLG}\panel{b} \\
    \bottomrule
    \end{booktabs}
    \caption{\figtitle{Summary of the correlated states observed at $B=0$.}
    }
    \label{table:CS}
\end{table*}

\end{document}